\documentclass[reprint, showpacs, superscriptaddress]{revtex4-1}
\usepackage[utf8]{inputenc}
\usepackage[english]{babel}
\usepackage{amsmath}
\usepackage{amsfonts}
\usepackage{amssymb}
\usepackage{graphicx}
\usepackage{dcolumn}
\usepackage{booktabs}
\usepackage{longtable}
\usepackage{afterpage}
\AtBeginDocument{
\heavyrulewidth=.08em
\lightrulewidth=.05em
\cmidrulewidth=.03em
\belowrulesep=.65ex
\belowbottomsep=0pt
\aboverulesep=.4ex
\abovetopsep=0pt
\cmidrulesep=\doublerulesep
\cmidrulekern=.5em
\defaultaddspace=.5em
} %This allows the use 'booktabs' rules
\begin{document}
\title{Theoretical calculation of atomic properties of superheavy elements $Z=110-112$ and their ions.}
\author{B.G.C. Lackenby}
\affiliation{School of Physics, University of New South Wales,  Sydney 2052,  Australia}
\author{V.A. Dzuba}
\affiliation{School of Physics, University of New South Wales,  Sydney 2052,  Australia}
\author{V.V. Flambaum}
\affiliation{School of Physics, University of New South Wales,  Sydney 2052,  Australia}
\affiliation{Johannes Gutenberg-Universit\"at Mainz, 55099 Mainz, Germany}

\begin{abstract}
We calculate the spectra, electric dipole transition rates and isotope shifts of the super heavy elements Ds ($Z=110$), Rg ($Z=111$) and Cn ($Z=112$) and  their ions. These calculations were performed using a recently developed, efficient version of the \textit{ab intio} configuration interaction combined with perturbation theory to treat distant effects. The successive ionization potentials of the three elements are also calculated and compared to lighter elements.
\end{abstract}
\maketitle

\section{Introduction}

The discovery and study of super heavy elements (SHEs) where $Z>103$ have been of great interest both experimentally and theoretically to physicists for the past 50 years. The large nuclear charge of these nuclei is predicted to result in exotic atomic properties which are not observed in other elements and breaks well established trends in the periodic table. While elements up to $Z=118$ have been experimentally synthesized and recognized, their low production rates and  short half lives have made the study of chemical and physical properties difficult  (see reviews \cite{Giuliani2019, schadel2015, HHO2013}). As such, there is no experimental data on their spectra though there has been experimental success in measuring the ionization potentials and single excited states in No and Lr which lie just below the SHEs\cite{Laatiaoui2016, Chhetri2018, SAB15}. Therefore, for the  further progress in experiment, the study of electron properties of SHEs must be made in the theoretical domain using many-body approaches. Such theoretical calculations will not only help us understand exotic properties SHEs, they are also predictive and will aid both future experimental measurements and the search for meta-stable isotopes which belong to a hypothetical island of stability  in astronomical data. \\
\linebreak
Most of SHE in the region $Z=103$ to $Z=118$ have open shells, with up to ten electrons in them. 
Theoretical study of such systems is difficult due to fast increase of the number of possible configurations with the number of electrons. There are few powerful methods of many-body relativistic calculations which work very well for atoms with relatively simple electron structure, having one to four electrons above closed shells. They were used in a number of studies of SHE which fell into this category (see, e.g.~\cite{eliav2015,pershina2015}). The use of these methods for systems with more than four external electrons is problematic due too high demand for computer power.
Also, techniques which extrapolate results of lighter elements are insufficient for treating SHEs due to the large relativistic effects which result in exotic properties which lighter elements in the elemental group do not have. An efficient method capable of calculating the spectroscopic properties of these elements has been developed in \cite{DBHF2017} which combines CI and perturbation theory (PT) referred to as the CIPT method.\\ 
\linebreak
The CIPT method has been used for the open $6d-$shell SHEs Db ($Z=105$)\cite{LDFDb2018}, Sg ($Z=106$), Bh($Z=107$), Hs($Z=108$) and Mt($Z=109$) \cite{LDFSg2019} along with the closed shell noble SHE Og ($Z=118$)\cite{LDFOg2018}. This method has also been used to accurately calculate the low-lying states of Ta\cite{LDFDb2018} and Rn\cite{LDFOg2018} (lighter elemental analogs of Db and Og respectively) when compared to available experimental data. This paper will focus on the heaviest SHE `metals' in groups 10, 11 and 12, specifically darmstadtium (Ds, $Z=110$), roentgenium (Rg, $Z=111$) and copernicium (Cn, $Z=112$). In particular, these SHEs are of interest as their proton number lies close to the expected magic number for stability, $Z=114$, and therefore the existence of long lived meta-stable isotopes is promising \cite{Giuliani2019}.  The expected magic numbers of neutrons for these meta-stable nuclei has been calculated to be $N=184$. The search for these meta-stable SHE and the ``island of stability'' has been at the frontier of nuclear physics for decades. These neutron rich nuclei cannot be produced in laboratory conditions, however it has been suggested that the neutron flux, which occurs in cosmological events, could create these meta-stable nuclei \cite{Goriely2011, Fuller2017, Friebel2018, Schuetrumpf2015}. A promising method for detecting traces of these isotopes in astrophysical data using calculated isotope shifts and experimental data from unstable, neutron deficient isotopes in laboratories was presented in Ref. \cite{DFW17}. Therefore, in this work we also present the isotope shifts of the neutral atoms and ions Ds I, II, III, and Rg I, II  for optical E1 transitions. While some states of Rg I and Cn I have been calculated previously (see Sections~\ref{sec:DsRg} and \ref{sec:Cn}) there has not been significant treatment of the ionic states of the elements or their isotope shifts. \\ 
\linebreak
The ground states for the three elements have been found to be [Rn]$6d^n7s^2$ where $n=8, 9, 10$ for Ds, Rg and Cn respectively. The Cn atom has relatively simple electron structure with completely closed shells in its ground state.
Therefore,  theoretical predictions of its spectra do exist, they were calculated using \textit{ab initio} techniques such as multiconfigurational Dirac Fock (MCDF)~\cite{Li2007, Yu2007}, relativistic pseudopotentials (RPP)~\cite{Hengele2012}, CI + MBPT~\cite{Dinh2008} and relativistic coupled cluster (RCC)~\cite{Eliav_Cn_1995} methods. Similarly for Rg, which is one electron short of a closed $6d$ shell, the RPP method has been used to calculate some states in the excitation spectrum with which we can compare our results.\\
\linebreak
This paper progresses as follows, in Section \ref{sec:CIPT} we give a brief overview of the CIPT method and how it is implemented for SHE. In Section \ref{sec:DsRg} and Section \ref{sec:Cn} we present the calculated low-lying excitation spectrum of Ds I, II, III, and Rg I, II and Cn I, II, III.  In Sections \ref{sec:E1IS} and \ref{sec:IP} we present the optical E1 transitions and corresponding isotope shifts, and the  successive ionization potentials of Ds, Rg and Cn respectively.

\section{The CIPT Method} \label{sec:CIPT}

As mentioned above, a novel configuration interaction approach to calculate the spectra of atoms with unfilled shells has been developed \cite{DBHF2017} and used to calculate the spectra, IPs and transition probabilities in SHEs $Z=102, 105-109$  and $Z=118$ and their respective lighter elemental analogs \cite{DBHF2017,LDFDb2018, LDFOg2018, LDFSg2019}. Recently the efficiency of this method has been improved upon with only a small additional cost of accuracy~\cite{FCI}. In this work we will give a brief overview of the method.\\
\linebreak
To generate the single-electron basis states a $V^{N_e-1}$ (where $N_e$ is the total number of electrons) Hartree-Fock (HF) approximation is used. In this approximation, the Hartree-Fock calculations are performed for the charged open-shell ion with one electron removed from the atom or ion of interest \cite{Kelly1964, Dzuba2005}. In most of cases external $s$ electron is removed. However, in some cases, like e.g. calculations for even states of Au~I and Hg~II, better accuracy is achieved if a $5d$ electron is removed. 
The multi-electron basis sets are then generated using a B-splines technique with 40 B-spline states in each partial wave of order 9 in a box of radius  $40 \ a_B$ (where $a_B$ is the Bohr radius) with partial waves up to $l_{\text{max}} = 4$. 
%Non relativistic reference configurations are chosen for the even and odd parity states of total angular momentum $J$, where  a new calculation is performed for each set of states with unique parity and $J$. 
The single determinant many-electron basis states $|i \rangle = \Phi_i(r_1,\dots,r_{N_e})$ for the CI calculations are  generated by making all single and double electron excitations from reference configurations.\\
\linebreak
The CI wavefunction, $| \Psi \rangle $, is written as an expansion over single-determinant many-electron states from two distinct sets of the many-electron basis states $\left|i\right>$, 
\begin{equation}
| \Psi \rangle = 
\sum_{i=1}^{N_{\text{Eff}}} c_{i}|i\rangle + \sum_{i = N_{\text{Eff}} + 1}^{N_{\text{total}}} c_{i}|i\rangle .
\label{eq:psi}
\end{equation}
The first summation in Eq. (\ref{eq:psi}) represents a small set of low energy wavefunctions which give a good approximation to the state ($i \leq N_{\text{Eff}}$, where $N_{\text{Eff}}$ is the number of wavefunctions in the low energy set). The second summation in Eq. (\ref{eq:psi}) is a large set of high energy-wavefunctions which are corrections to the state. The CI matrix is constructed by ordering the basis states $\left|i\right>$ by energy and divided into the two sets of low energy and high energy states. The CI matrix is simplified by neglecting all off diagonal matrix elements of  the CI matrix between the terms in the high energy set, $\langle i | H^{\text{CI}} | j \rangle = 0 $ for $|i\rangle, |j\rangle > N_{\text{Eff}}$. This truncation of the matrix significantly reduces the previous large diagonalization problem to a simplified CI diagonalization problem of size $N_{\text{Eff}}$. This smaller matrix of size $N_{\text{Eff}} \times N_{\text{Eff}}$ is referred to as the effective CI matrix.
\begin{equation} \label{eq:CI}
(H^{\rm CI} - EI)X=0,
\end{equation}
where $I$ is unit matrix, the vector $X = \{c_1, \dots, c_{N_{\rm eff}}\}$. The high energy basis states $i > N_{\text{Eff}}$ are included  by modifying the matrix elements of the effective CI matrix. Specifically the matrix elements of the effective CI matrix are modified to include perturbative contributions from the high energy states,
\begin{equation}
\langle i|H^{\rm CI}|j\rangle \rightarrow \langle i|H^{\rm CI}|j\rangle + 
\sum_k \frac{\langle i|H^{\rm CI}|k\rangle\langle k|H^{\rm
    CI}|j\rangle}{E - E_k}. 
    \label{eq:HCI}
\end{equation}
where $i , j \leq N_{\text{Eff}}$, $k > N_{\text{Eff}}$,  $E_k = \langle k|H^{\rm CI}|k\rangle$, and $E$ is the energy of the state of interest. As this energy is not known \textit{a priori}, iterations of the second summation must be performed until there is a convergence in $E$. When this convergence is achieved, the energy is an exact solution to the truncated CI matrix. 
%The perturbation does not introduce any loss of accuracy. 
This is known as the CIPT method. \\
\linebreak
The Breit interaction \cite{Breit1929, Mann1971}  and quantum electrodynamic (QED) radiative corrections  (Ueling potential and electric and magnetic form factors) \cite{FG2005} are included in the calculations as described in our earlier works (see, e.g. \cite{FF113-115}).  As both the Breit and QED radiative corrections scale with nuclear charge, $Z$, faster than the first power \cite{FF113-115} their contribution to the energy levels of SHE is non negligible. For each level we calculate the Land\'{e} $g$-factor for comparison with experimental $g$-factors in lighter elements. To label the levels in the SHE spectra for reference, we compare the SHE states to similar states in lighter analogs with similar $g-$factors, and if available, adopt that notation for the SHE state. However it should be noted that $LS$ notation is not, in general, appropriate for labeling SHE states. This is 
%due to the large relativistic effects 
due to very large spin-orbit interaction in SHEs (so the eigenvectors will look strongly mixed in $LS$ notation). We only use $LS$ notations for comparison with lighter elements. If analogous states of the lighter element are not available with LS notation we label the $n$th sequential state of total angular momentum $J$ and parity by $n_{J}^{\text{parity}}$. We use the same notation for presenting states of lighter elements when $LS$ notation is not available.

\section{Ds and Rg} \label{sec:DsRg}

Both elements darmstadtium and roentgenium  were first synthesized in 1994 \cite{Ds1994, Rg1994} and officially named and recognized in 2001 \cite{DsIUPAC}. Early theoretical calculations of their ground states show that they are anomalous in each of their groups. Consider both of the lighter elemental analogs Pt and Au which have ground states $5d^96s$ and $5d^{10}6s$ respectively. The large relativistic effects in Ds and Rg directly stabilize the 7s orbital and indirectly destabilize the $6d$ orbital resulting in ground states of $6d^{8}7s^2$ and $6d^97s^2$\cite{Eliav_Rg_1994} respectively. The ground state of Ds and Rg both follow the same trend of the other open $6d$-shell elements which all have closed $7s$ shell ground states. Using the CIPT method described in Section \ref{sec:CIPT} we calculate the low-lying excitation spectrum of   both Ds I, II, III in  Table \ref{tab:PtDsspectrum} and Rg I, II in Table \ref{tab:Rgspectrum}. To gauge the accuracy of the atomic calculations we also calculated the energy levels of Pt I, Pt II and Pd III for comparison with available experimental results in Table \ref{tab:PtDsspectrum}.  
\begin{longtable*}{l@{\hspace{0.5cm}}cr@{\hspace{0.5cm}}r@{\hspace{0.5cm}}r@{\hspace{0.5cm}}r@{\hspace{0.5cm}}r@{\hspace{0.5cm}}|l@{\hspace{0.5cm}}cr@{\hspace{0.5cm}}r@{\hspace{0.5cm}}r@{\hspace{0.5cm}}r}
\caption{Comparison of CIPT energy level calculations for neutral Pt I, II and Pd III with experimental values. Low lying even and odd states for Ds I, Ds II and Ds III calculated using the CIPT energy. Experimental energies and CIPT energies are given by $E_{\text{E}}$ and $E_{\text{T}}$ respectively. Where available, the experimental Land\'{e} g-factors $g_{\text{E}}$ are provided for comparison. The discrepancy between the experimental and CIPT energies is given by $\Delta = E_{\text{E}} - E_{\text{T}}$.\label{tab:PtDsspectrum}}
\endfirsthead
\toprule
\toprule
 &           &  \multicolumn{2}{c}{Experimental\cite{NIST_ASD}} & \multicolumn{2}{c}{CIPT}  & & &  &\multicolumn{2}{c}{CIPT} \\
 \cline{3-4} \cline{5-6} \cline{10-11}
 \\
& State  & \multicolumn{1}{c}{\parbox{1cm}{$E_E$ \\ (cm$^{-1}$)}} & $g_{\text{E}}$  & \multicolumn{1}{c}{\parbox{1cm}{$E_T$ \\ (cm$^{-1}$)}}& $g_{\text{T}}$ & $\Delta$ & & State & $E_{\text{T}}$ & $g_{\text{T}}$  \\
 \midrule
\endhead
\hline \multicolumn{5}{r}{\textit{Continued on next page}} \\
\endfoot
\endlastfoot
\toprule
\toprule
 &           &  \multicolumn{2}{c}{Experimental\cite{NIST_ASD}} & \multicolumn{2}{c}{CIPT} & & &  &\multicolumn{2}{c}{CIPT} \\
 \cline{3-4} \cline{5-6} \cline{10-11}
 \\
& State  & \multicolumn{1}{c}{\parbox{1cm}{$E_E$ \\ (cm$^{-1}$)}} & $g_{\text{E}}$  & \multicolumn{1}{c}{\parbox{1cm}{$E_T$ \\ (cm$^{-1}$)}}& $g_{\text{T}}$ & $\Delta$ & & State & $E_{\text{T}}$ & $g_{\text{T}}$  \\
 \midrule
\toprule
 \multicolumn{7}{c}{Pt  I} &  \multicolumn{4}{c}{Ds I}\\
 \multicolumn{6}{l}{\textit{Even states}} & & & &\\
$5d^96s$ & $^3$D$_3$ & 0 &  & 0  & 1.33 &  &  $6d^8 7s^2$ &   $^3$F$_4$   &  0 &   1.23     \\
$5d^96s$ & $^3$D$_2$  & 776 & 1.01 & 728  &    1.07   & 48  & $6d^8 7s^2$ &   $^3$P$_2$   & 4 146 &  1.12\\
$5d^86s^2$ & $^3$F$_4$  & 824 & 1.25 & 1 289 &1.24         & -465  &     $6d^8 7s^2$   &    0    &   14 541 &  0.00  \\
$5d^{10}$ & $^1$S$_0$ &  6 140 & &  5 148   & 0.00    &  992 & $6d^8 7s^2$ &    $^3$F$_3$  & 16 499 &   1.08 \\
$5d^96s$ & $^3$D$_1$	  &	  10 132 &   	&  8 889  & 0.50     & 1 243 & $6d^8 7s^2$    & $^3$P$_1$  &    23 322 & 1.50   \\
 \multicolumn{6}{l}{\textit{Odd states}} & & & &\\
$5d^86s6p$  	&  $^5$D$^{\rm o}_4$  	&  30 157    	 &  	  1.46  &    31 390 &  1.46   & -1233   &  $6d^7 7s^2 7p$  & 1$^{\rm o}_4$ &  21 812 &   1.34      \\
$5d^96p$  	 &  	 1$^{\rm o}_2$  	&  32 620     	  & 	  1.39  	   &  31 652    &  1.38     &  968 	&   $6d^7 7s^2 7p$ & 1$^{\rm o}_5$ & 24 958 &   1.23   \\
$5d^86s6p$  	 & 	 1$^{\rm o}_5$  &	  33 681  &	   	  1.32  	  &    34 662    & 1.31   & -981 &    $6d^7 7s^2 7p$ & 1$^{\rm o}_1$ &   26 779 & 1.44   \\
$5d^96p$  	  & 	 1$^{\rm o}_3$  	&  34 122    	&  	  1.21  	 &  33 141    & 1.13   & 981 &   $6d^7 7s^2 7p$ & 1$^{\rm o}_2$ &  28 550 &   1.07      \\   	 
$5d^86s6p$  	  &	2$^{\rm o}_3$  &	  35 322  &  	   	  1.33  &   36 479  &  1.36   & 	-1157 &    $6d^7 7s^2 7p$ & 2$^{\rm o}_2$ &    30 383 & 1.34   \\
$5d^86s6p$  	 & 	 2$^{\rm o}_4$  &	  36 296    & &  36 394  &  1.25     &   -98   &  $6d^7 7s^2 7p$ & 1$^{\rm o}_3$ &   32 645 &  1.16	  \\
$5d^86s6p$ 	&  	 $^5$G$_6$  &	  36 782     	&   	  1.33  	& 37 603  & 1.33 &	-821     &    $6d^7 7s^2 7p$ & 2$^{\rm o}_3$ &  36 404  & 1.25    \\
$5d^96p$  	 & 	 2$^{\rm o}_1$ & 	  36 845     &	   	  1.09  	 & 36 761 & 1.16       & 84  	& $6d^7 7s^2 7p$ & 2$^{\rm o}_4$ & 34 919  & 1.38   \\
$5d^96p$  	 & 	 1$^{\rm o}_2$ &	  37 342     	&   	  1.15  	   &  36 889   & 1.14   & 453  &    $6d^7 7s^2 7p$ & 3$^{\rm o}_4$ &   39 814 & 1.15   \\ 	      
$5d^96p$  	& 	 3$^{\rm o}_4$  	&  37 591    &	   	  1.25  	  &  37 615    & 1.17      & 	-24   	 &   	 $6d^7 7s^2 7p$ & 2$^{\rm o}_5$ &   40 173 &     1.27  \\ 	 	  
$5d^96p$   	&  	 3$^{\rm o}_3$  	&  37 769     	&   	  1.17  	  &   37 218   & 1.24 &    551    &  $6d^7 7s^2 7p$ & 1$^{\rm o}_0$ & 40 668   &  0.00   \\ 	  	  	  	  	  	  	  	  	  	 
$5d^86s6p$ 	 &  	 $^5$F$^{\rm o}_5$  	&  38 536    & 	   	  1.30  	 &  39 451    & 1.31       &  -915    & $6d^7 7s^2 7p$ & 3$^{\rm o}_2$ & 42 632 &   1.29  \\	  	  	  	 
$5d^86s6p$  	 &	 2$^{\rm o}_2$  	&  38 816     	 &  	  0.88  &   39 275   & 0.76     & -459	 &     $6d^7 7s^2 7p $   &   3$_5^{\rm o}$  &   42 682  & 1.22   \\ 	  	  	  	 
$5d^86s6p$  	 &	 4$^{\rm o}_4$  	&  40 194.2     	 &  	  1.21  &   41 329     & 1.23    &  -1 135  &    $6d^7 7s^2 7p $ &  3$_3^{\rm o}$    &       42 722 &  1.26      \\ 	 
$5d^86s6p$  	 &	3$^{\rm o}_2$  	&  40 516.3    	 &  	  1.38  &  41 968    & 1.23   & -1 452 &  $6d^7 7s^2 7p$    &    1$_6$   &   42 322	& 1.28   \\ 	 	  	  
$5d^86s6p$  	 &	 4$^{\rm o}_2$  	&   40 787.9     	 &  	  1.20  &   42 262  &1.35         &  -1 474  &       $6d^7 7s^2 7p $ &  2$_1^{\rm o}$ &      42 828  &  0.78     \\ 	 	  	  	  	  
$5d^96p$  	 &	 2$^{\rm o}_0$  	&  40 873.5     	 &  	    &    41 467   &  0.00     & -594 &     $6d^7 7s^2 7p $  & 4$_2^{\rm o}$ &      42 900  &  1.10     \\ 	  	  	  	  	  
$5d^86s6p$  	 &	 4$^{\rm o}_3$  	&   40 970.1     	 &  	  1.12  &  41 991   & 1.09   & -1 021 &    $6d^7 7s^2 7p$ & 4$_4^{\rm o}$   &       43 915 & 1.18   \\ 	 
$5d^86s6p$  	 &	 3$^{\rm o}_1$  	&  41 802.7    	 &  	  0.92  &  41 916  &  0.82     &  -113 &   $6d^7 7s^2 7p $ &  4$_3^{\rm o}$   &         44 974  &1.23       \\ 	 
$5d^86s6p$  	 &	 5$^{\rm o}_3$  	&  42 660.2    	 &  	   1.19  &  44 087   & 1.14     & -1 427  &   $6d^8 7s  7p $  & 5$_2^{\rm o}$ &           45 149   &	1.29       \\
$5d^86s6p$  	 &	 4$^{\rm o}_1$  	&  43 187.8     	 &  	  1.39  &  44 300 &  1.31     & -1 112 &    $6d^7 7s^2 7p $  & 2$_0^{\rm o}$  &      46 009  &  0.00         \\
\\
\midrule
								 \multicolumn{7}{c}{Pt II} & \multicolumn{4}{c}{Ds II}  \\
								  \multicolumn{6}{l}{\textit{Even states}} & & & &\\
$5d^9$  	& $^{2}$D$_{5/2} $   &   	   	  0.0  &   &  0 & 1.20 &  &   $6d^7 7s^2$ & $^4$F$_{9/2}$ & 0 & 1.27\\
$5d^86s$  	&$^{4}$F$_{9/2} $  &    	   	  4 786.6    &   & 4 653  & 1.33    & 134  &  $6d^7 7s^2$	 & $^4$F$_{3/2}$  	& 4 464  & 1.20    \\
$5d^9$  &	$^{2}$D$_{3/2} $   &   	   	  8 419.9     	&    & 8 031 & 0.79   & 389 &  $6d^7 7s^2$	& $^4$F$_{5/2}$ & 8 484 & 1.21  \\
$5d^86s$ &$^{4}$F$_{7/2} $	&      	   	  9 356.2    &   	 &  9 166 & 1.20  	 & 190 &  $6d^7 7s^2$ & $^4$F$_{7/2}$ & 15 407 &  1.20  \\
$5d^86s$ & $^4$P$_{1/2}$ & 21 718 &  & 22 886 & 2.57 & -1168 &  $6d^7 7s^2$ & 1$_{1/2}$ & 21 178 & 1.37 \\
& & & & & & &     $6d^7 7s^2$ & 1$_{11/2}$ & 28 183 & 1.09	\\
 \multicolumn{6}{l}{\textit{Odd states}} & & & & \\
 $5d^86p$ & $^4$D$_{7/2}^{\rm o}$ &  51 408 &  & 52 054 & 1.35 & -646  & $6d^6 7s^2 7p$ & 1$_{7/2}^{\rm o}$ & 37 951 & 1.43 & \\
$5d^86p$ & $^4$G$_{9/2}^{\rm o}$ &  53 876 &  & 54 046 & 1.18 & -170 & $6d^6 7s^2 7p$ & 1$_{3/2}^{\rm o}$ & 38 415 & 1.48 \\
$5d^86p$ & $^4$D$_{3/2}^{\rm o}$ & 56 588 &  & 58 285& 1.26 & -1697 & $6d^7 7s 7p$ & 1$_{9/2}^{\rm o}$ & 39 010 & 1.42 \\
$5d^86p$ & $^4$G$_{5/2}^{\rm o}$ & 57 018 &  &58 198 & 1.16 & -1180 & $6d^6 7s^2 7p$ & 1$_{5/2}^{\rm o}$ & 40 654 & 1.25 \\
& & & & & & & $6d^6 7s^2 7p$ & 2$_{9/2}^{\rm o}$ & 41 911 & 1.28 \\
\midrule
 \multicolumn{7}{c}{Pd III} &  \multicolumn{4}{c}{Ds III}\\
 								  \multicolumn{6}{l}{\textit{Even states}} & & & & \\
   $4d^8$ &$^3$F$_4$ & 0 & & 0  & 1.25     & &          $6d^6 7s^2$  &  1$_2$   &       0  & 1.37   \\
   $4d^8$ & $^3$F$_3$ & 3 229 & & 3 173 & 1.08    &56 &          $6d^7 7s $ &    $^5$F$_5$   &      1 412 &   1.35        \\
  $4d^8$ & $^3$F$_2$  & 4 686 & & 4 695 & 0.70   & -9 &    $6d^6 7s^2$ &  $^5$F$_4$    &     1 675 & 1.32      \\
   $4d^8$ & $^3$P$_1$ & 13 469 &  & 14 787 & 1.50   & -1 318 &           $6d^6 7s^2$ &  1$_0$ &    12 795   &     0.00     \\
   $4d^8$ &  $^3$P$_0$ & 13 698 & & 15 212& 0.00  & -1 514 &        $6d^7 7s $ &    $^5$F$_1$      & 13 279   & 1.15     \\
   $4d^7 5s$ & $^5$F$_5$  & 52 916 &  &  51 230 &  1.40 & 1 686 &       $6d^7 7s $ &  $^5$F$_3$   &     13 376  &  1.38     \\
    \multicolumn{6}{l}{\textit{Odd states}} & & & & \\
 $4d^7 5p$ & $^5$D$_4^{\rm o}$ & 104 418.86 & &101 793 &   1.41 & -2 626 &    $6d^6 7s  7p $ & 1$_4^{\rm o}$   &     45 317   &   1.54    \\
&   & &  &  & &    &$6d^6 7s  7p $  &    1$_2^{\rm o}$  &      48 521   & 1.71  \\
&   & &  &  & & &$6d^6 7s  7p $   &  1$_5^{\rm o}$ &      51 498   & 1.41    \\
\bottomrule
\bottomrule
\end{longtable*}

For Pt I and Pt II the even states are calculated using the reference states $5d^{n-1}6s$, $ 5d^{n-2}6s^2$ and $5d^{n}$  and odd states are calculated with reference states $5d^{n-2}6s6p$, $5d^{n-3}6s^26p$ and $5d^{n-1}6p$ where $n=10$ and $9$ for the neutral atom and ion respectively. The CI matrix is populated with all single and double excitations of these reference states. Similarly for the calculations of Pd III we used reference states $4d^{7}5s$, $ 4d^{6}6s^2$ and $4d^{8}$ for the even states and $4d^85s5p$, $4d^75s^25p$ and $4d^95p$ for odd states. The spectrum of Pd III is calculated for comparison as there is no available experimental data for Pt III.\\
\linebreak
From Table \ref{tab:PtDsspectrum} we see that there is good agreement between experimental results aggregated in Ref. \cite{NIST_ASD} and the CIPT calculations of Pt I, II and Pd III. While not as consistently accurate as the calculations of Ta in Ref. \cite{LDFDb2018} which had an accuracy of $|\Delta| \approx 500$~cm$^{-1}$, for both the odd and even parity states of Pt I and Pt II there is agreement to within $|\Delta| \approx 1500$ ~cm$^{-1}$for low-lying states. For higher states the absolute energy difference between experimental and theoretical results is larger but the relative difference is only $\sim$ 2\%.
We expect this level of accuracy to be similar for the calculations of neutral Ds and the respective ions using the CIPT method. \\
\linebreak
To calculate the excitation spectrum of Ds I, II and III reference configurations $6d^{n-1}7s$, $ 6d^{n-2}7s^2$ and $6d^{n}$ (even states) and, $6d^{n-2}7s7p$, $6d^{n-3}7s^27p$ and $6d^{n-1}7p$ (odd states) are used to populate the effective CI matrix for $n=10, 9$ and $8$ respectively. Comparing the spectra of neutral Ds and its ions with the spectra of lighter elemental analog we see that while they are in the same elemental group in the periodic table, there are some stark differences between their spectra. As previously mentioned, the ground state of Ds I follows the SHE trend of a closed $7s^2$ shell unlike Pt I. The relativistic contraction of the $7s$ shell and consequent destabilization of the $6d$ shell in the SHE spectra results in a majority of odd parity states from the excitation of the $6d$ electron to the $7p$ shell. Comparatively, the lighter analog spectrum of odd states is dominated by excitations of the $6s$ electron to the $6p$ shell. This lowers the odd state spectrum of Ds I with the first odd parity state 1$^{\rm o}_4$ at 21812~cm$^{-1}$ compared to the lowest state  $^5$D$^{\rm o}_4$  	at  30157~cm$^{-1}$.  This can also be seen in the odd parity spectrum of Ds II when compared to Pt II. In Pt II the first odd parity state is located far outside the optical region at 51408~cm$^{-1}$ while  there are at least 5 odd parity states in the Ds II spectrum which could potentially  be detected through optical transitions to the ground state. This is similar to what was found when comparing the lighter open $6d-$shell elements to their respectively lighter analogs in Refs. \cite{LDFDb2018, LDFSg2019}. The electric dipole (E1) amplitudes and transition rates of these optically accessible states are calculate in Section \ref{sec:E1IS}.\\
\linebreak
\begin{table}
\caption{Comparison of CIPT energy level calculations, $E_{\text{T}}$ and experimental energy levels, $E_{\text{E}}$,results for Au and Au II. Where available, the experimental Land\'{e} $g$-factors, $g_{\text{E}}$, are given along with calculated $g$-factors, $g_{\text{T}}$. The difference between the experimental and theoretical energies are also presented, $\Delta = E_{\text{E}} - E_{\text{T}}$.   \label{tab:AuComp}}
\begin{tabular}{l@{\hspace{0.2cm}}cr@{\hspace{0.2cm}}r@{\hspace{0.2cm}}r@{\hspace{0.2cm}}r@{\hspace{0.2cm}}r@{\hspace{0.2cm}}}
\toprule
\toprule
& State  & $E_{\text{E}}$ \cite{NIST_ASD}& $g_{\text{E}}$  & $E_{\text{T}}$ & $g_{\text{T}}$ & $\Delta$\\
 &           & (cm$^{-1}$)            &                 &   (cm$^{-1}$) & (cm$^{-1}$) & (cm$^{-1}$)  \\
 \midrule
 \multicolumn{7}{c}{Au I} \\
  \multicolumn{7}{l}{\textit{Even states}}\\
$5d^{10}6s$ &  $^2$S$_{1/2}$  & 0 & 1.997 & 0  & 2.00  & \\
$5d^{9}6s^2$ &  $^2$D$_{5/2}$  & 9 161.77  & 1.192 & 10 902 & 1.20 & -1 740 \\
$5d^{9}6s^2$ &  $^2$D$_{3/2}$  & 21 435.191 & 0.804 & 22 361  & 0.80  & -926 \\
\multicolumn{7}{l}{\textit{Odd states}} \\
$5d^{10}6p$ &  $^2$P$_{1/2}^{\rm o}$ 	  & 37 358.991  &  0.661   & 38 722 & 0.67  & -1 363   \\
$5d^{10}6p$ 	&  $^2$P$_{3/2}^{\rm o}$   &  41 174.613  & 1.334   & 42 648 & 1.33 & -1 473  \\
\midrule
								 \multicolumn{7}{c}{Au II} \\
								   \multicolumn{7}{l}{\textit{Even states}}\\
$5d^{10}$ &    $^1$S$_0$   &0  & & 0 & 0 & \\
$ 5d^9 6s$ &  $^3$D$_3$   & 15 039.572 &  & 15 887  & 1.33  & -847  \\
$ 5d^9 6s$ & $^3$D$_2$   &   17 640.616 & & 18 551 & 1.20 &  -910 \\
$ 5d^9 6s$ & $^3$D$_1$  & 27 765.758  & & 27 854 & 0.50 &  -88 \\
\multicolumn{7}{l}{\textit{Odd states}} \\
 $5d^{9}6p$ &   1$_2^{\rm o}$  & 63 053.318  & 1.45 & 64 964 & 1.39   & -1 910  \\
\bottomrule
\bottomrule
\end{tabular}
\end{table}
The excitation spectrum of Rg I and Rg II was calculated using the CIPT method and the results are presented in Table \ref{tab:Rgspectrum}. The CIPT calculation of Rg was very similar to Ds. The reference configurations used to populate the CI matrix are $6d^{n-1}7s$, $ 6d^{n-2}7s^2$  (even states) and, $6d^{n-2}7s7p$, $6d^{n-3}7s^27p$ and $6d^{n-1}7p$ (odd states) where $n=11$ and $10$ for Rg I and Rg II respectively. As for the Pt/Pd and Ds calculations both the neutral and first ion spectrum of Au were calculated to determine the accuracy of the Rg calculations. The Au calculations used appropriate analogous reference configurations to those used for Rg and are presented in Table \ref{tab:AuComp} with experimental results for comparison. We see that the accuracy of the Au calculations is similar to that for Pt and Pd ($|\Delta| \approx 1500$ ~cm$^{-1}$) . We expect a similar accuracy for the CIPT calculations of Rg I and II are presented in Table \ref{tab:Rgspectrum}. These results agree with the early calculations of Ref. \cite{Eliav_Rg_1994} which found Rg I has a ground state of $6d^97s^2$ ($^2$D$_{5/2}$). As seen in Ds, the odd parity energies of Rg I have been shifted lower in the spectrum when compared to Au I.  In Au I there are at most 2 viable optical E1 transitions from the ground state whereas in Rg I there are 5 promising transitions. In Au II there are no optically accessible states where there two potential states in Rg II.\\
\linebreak
\begin{table}
\caption{Low lying even and odd states for Rg I and Rg II calculated using the CIPT energy. The theoretical CIPT energies are given by $E_{\text{CIPT}}$ and the Land\'{e} g-factors are given by $g_{\text{CIPT}}$. Where available, previously calculated states in Ref. \cite{Hengele2012} are given by $E_{\text{PP}}$ for comparison. \label{tab:Rgspectrum}}
\begin{tabular}{l@{\hspace{0.3cm}}cr@{\hspace{0.3cm}}r@{\hspace{0.3cm}}r@{\hspace{0.3cm}}r@{\hspace{0.3cm}}r}
\toprule
\toprule
& State  & $E_{\text{CIPT}}$  & $g_{\text{CIPT}}$ & $E_{\text{RCC}}$ \cite{Eliav_Rg_1994} & $E_{\text{PP}}$ \cite{Hengele2012}  \\
 &           & (cm$^{-1}$)            &                 &   (cm$^{-1}$) & (cm$^{-1}$)  \\
 \midrule
\multicolumn{6}{c}{Rg I}\\
  \multicolumn{6}{l}{\textit{Even states}}\\
$6d^9 7s^2$  & $^2$D$_{5/2}$  &      0  &  1.20   \\
$6d^9 7s^2$ & $^2$D$_{3/2}$ &      19 174  & 0.80  & 21 670 & 20 250   \\
$6d^{10} 7s$ & $^2$S$_{1/2}$  &   22 428  &   2.00  & 23 820 & 24 760   \\
\multicolumn{6}{l}{\textit{Odd states}} \\
$6d^8 7s^2 7p$  &  1$_{7/2}^{\rm o}$   &        28 224 & 1.32    \\
$6d^8 7s^2 7p$  &  1$_{9/2}^{\rm o}$    &   31 795 &    1.17   \\
$6d^8 7s^2 7p$  &  1$_{3/2}^{\rm o}$  &     32 677  & 1.13      \\
$6d^8 7s^2 7p$ &   1$_{5/2}^{\rm o}$    &     34 398  & 1.07      \\
$6d^9 7s  7p$ & $^4$P$_{5/2}^{\rm o}$     &        42 709 &  1.46   \\
$6d^8 7s^2 7p$  &   1$_{1/2}^{\rm o}$    &       44 292 & 0.72  \\
$6d^9 7s  7p$   &  $^4$F$_{7/2}^{\rm o}$     &      46 619 &1.22    \\
$6d^8 7s^2 7p$ &    3$_{5/2}^{\rm o}$  &            47 517 & 1.13  \\
$6d^9 7s  7p$ & $^4$P$_{3/2}^{\rm o}$  &     48 547  &  1.40       \\
\\
\midrule
\multicolumn{6}{c}{Rg II} \\
  \multicolumn{6}{l}{\textit{Even states}}\\
       $6d^8 7s^2$  &   $^3$F$_4$         &   0 & 1.23  \\
        $6d^8 7s^2$ &  1$_2$         &   3 786 & 1.11  \\
         $6d^9 7s$   & $^3$D$_3$      &   12 255 & 1.33  &  13 950 & 16 720 \\
        $6d^8 7s^2$   &  1$_0$    &   15 754 &    0.00 \\
         $6d^8 7s^2$ &  $^3$P$_1$    &  28 105  &  1.50    \\
         \multicolumn{6}{l}{\textit{Odd states}} \\
     $6d^7 7s^2 7p$    & 1$_4^{\rm o}$       &  42 047   & 1.32 \\
      $6d^8 7s  7p$  &  2$_4^{\rm o}$   &  44 863 & 1.41    \\
     $6d^7 7s^2 7p$  &  1$_1^{\rm o}$        &   45 219  & 1.42 \\
    $6d^7 7s^2 7p$   &    1$_5^{\rm o}$      &   45 926 &  1.23 \\
    $6d^7 7s^2 7p$   &  1$_2^{\rm o}$        & 47 132  &  1.25  \\
    $6d^7 7s^2 7p$   &    2$_2^{\rm o}$      & 47 915   & 1.14  \\
   
\bottomrule
\bottomrule
\end{tabular}
\end{table}
Unlike Ds, some excitation levels and ionization potentials of Rg I have been previously calculated in Ref. \cite{Hengele2012} using a pseudo-potential method and in Ref. \cite{Eliav_Rg_1994} using a relativistic coupled cluster method. These values are included for comparison in Table \ref{tab:Rgspectrum}. While our CIPT calculations are in good agreement with these calculations, they are always lower. While there has been calculation of odd parity states of Rg I in Ref. \cite{Hengele2012} they consider the excitation $7s \rightarrow 7p$ above a closed $6d$ shell.  The E1 transitions of Rg I and Rg II along with the corresponding isotope shifts have been included in  Section~\ref{sec:E1IS}.

\section{Cn } \label{sec:Cn}

Copernicium (Cn) was first synthesized in 1996 \cite{Hofmann1996} in Darmstadt Germany. In particular the isotope $^{277}$Cn was synthesized which has a halflife of $200$~ps which is too short for chemical study. Compared to Ds and Rg, there has been considerably more theoretical and experimental study on Cn where chemical properties such as its the interaction with gold have been investigated \cite{Eichler2007}. This is primarily due to the closed $6d$ shell in the ground and some excited states of Cn. Calculations for such states can be done with many different methods. There has been significant theoretical study on the excitation spectrum compared to the lighter SHE. Many body techniques such as relativistic coupled cluster (RCC) \cite{Eliav_Cn_1995}, Multiconfigurational Dirac Fock (MCDF) \cite{Li2007, Yu2007}, relativistic Hartree-Fock and CI \cite{Dinh2008} and relativistic pseudo-potentials \cite{Hengele2012} have been used to calculate the exciation energies, ionisation potentials and oscillator strengths of Cn. Unlike the other SHE there has also been studies on the first and second ions of Cn in \cite{Eliav_Cn_1995, Yu2007}.  Using the CIPT method we compared our calculations of neutral Hg and the ions Hg II and Hg III with experimental results, the results are presented in Table \ref{tab:HgComp}. There are only few low energy states in the excitation spectrum of Hg I, II and III due to the stability of the closed shells. We find good agreement between the experimental and CIPT results of $|\Delta| < 1000$ cm$^{-1}$ in Table \ref{tab:HgComp}. 
\begin{table}
\caption{Comparison of CIPT energy level calculations, $E_{\text{T}}$, and experimental energy levels, $E_{\text{E}}$,results for neutral Hg and ions. Where available, the experimental Land\'{e} $g$-factors, $g_{\text{E}}$, are given along with calculated $g$-factors, $g_{\text{T}}$. The difference between the experimental and theoretical energies are also presented, $\Delta = E_{\text{E}} - E_{\text{T}}$. \label{tab:HgComp}}
\begin{tabular}{l@{\hspace{0.2cm}}cr@{\hspace{0.2cm}}r@{\hspace{0.2cm}}r@{\hspace{0.2cm}}r@{\hspace{0.2cm}}r@{\hspace{0.2cm}}}
\toprule
\toprule
& State  & $E_{\text{E}}$ \cite{NIST_ASD} & $g_{\text{E}}$  & $E_{\text{T}}$ & $g_{\text{T}}$ & $\Delta$\\
 &           & (cm$^{-1}$)            &                 &   (cm$^{-1}$) & (cm$^{-1}$) & (cm$^{-1}$)  \\
 \midrule
 \multicolumn{7}{c}{Hg I} \\
$5d^{10}6s^2$ & $^1$S$_0$   & 0 &  & 0 & 0 & \\
$5d^{10}6s6p$ & $^3$P$^{\rm o}_0$    & 37 645 &  & 37 572 & 0 & 73  \\
$5d^{10}6s6p$ 	&  $^3$P$^{\rm o}_1$   & 39 412 &  1.48 & 39 124 & 1.49 & 288  \\
$5d^{10}6s6p$	& $^3$P$^{\rm o}_2$   & 44 043 & 1.50 & 43 623 & 1.50  & 420  \\
$5d^{10}6s6p$ & $^3$P$^{\rm o}_1$    & 54 068 &  & 52 658   & 1.02 & 1 410  \\
\midrule
								 \multicolumn{7}{c}{Hg II} \\
$5d^{10}6s$ & $^2$S$_{1/2}$   & 0.00 & & 0.00 & 2.00 & \\
$ 5d^9 6s^2$ & $^2$D$_{5/2}$  & 35 515 & & 37 278 & 1.20 & -1 763  \\
 $5d^{10}6p$ & $^2$P$^{\rm o}_{1/2}$   & 51 486 & & 52 130 & 0.67  & -644\\
$5d^9 6s^2$ &  $^2$D$_{3/2}$ & 50 556 & &  51 423 & 0.80 & -867  \\
 $5d^{10}6p$ & $^2$P$^{\rm o}_{3/2}$   & 60 608 & & 60 860 & 1.33 & -252\\
 \midrule
 								 \multicolumn{7}{c}{Hg III} \\
$5d^{10}$  &	 $^1$S$_0$  &	   	  0.0  & & 0 \\  	   	
$5d^{9}6s$  &	 	 1$_3$  	&  42 850.3  &   & 43 791 & 1.33 & -941 \\
$5d^{9}6s$  &	  	 1$_2$  &	  46 029.5  &   &	 46 997 & 1.17 & -968  \\ 	   	
$5d^{9}6s$  &	  	 1$_1$  	&  58 405.8  &  & 58 538 & 0.50 & -132 	\\
\bottomrule
\bottomrule
\end{tabular}
\end{table}
In Table \ref{tab:CnSpectrum} the low-lying spectrum of Cn is presented and compared to other calculations. 
\begin{table}
\caption{Comparison of theoretical energy level calculations for neutral Cn. The energy levels of this work are given by $E_{\text{CIPT}}$ with $g-$factors $g_{\text{CIPT}}$. Where available, previous atomic calculations using multiconfigurational Dirac-Fock ($E_{\text{MCDF}}$) and relativistic Hartree-Fock calculations ($E_{\text{RHF}}$) are presented for comparison.  \label{tab:CnSpectrum}}
\begin{tabular}{l@{\hspace{0.3cm}}c@{\hspace{0.3cm}}r@{\hspace{0.3cm}}r@{\hspace{0.3cm}}r@{\hspace{0.3cm}}r@{\hspace{0.3cm}}r}
\toprule
\toprule
& State  & $E_{\text{CIPT}}$ & $g_{\text{CIPT}}$  & $E_{\text{RHF}}$ \cite{Dinh2008} & $E_{\text{MCDF}}$ \cite{Li2007} \\
 &           & (cm$^{-1}$)            &                 &   (cm$^{-1}$) & (cm$^{-1}$) \\
 \midrule
$6d^{10}7s^2$  & $^1$S$_0$ & 0 & 0 & 0 & 0   \\
$6d^{9}7s^27p$ & 1$_2$  & 31 263& 1.37 & 35 785  & 34 150   \\
$6d^{9}7s^27p$ & 1$_3$  & 33 857 & 1.10 & 38 625 & 37 642    \\
$6d^{10} 7s 7p$    & $^3$P$_0^{\rm o}$  &  45 097 &   0.00 & 51 212 & 48 471 &     \\
$6d^{10} 7s 7p$ &$^3$P$_1^{\rm o}$  & 47 293 &   1.41 & 53 144  & 52 024  &  \\
$6d^9 7s^2 7p$  & $^3$P$_2^{\rm o}$ &  54 241  &  0.98 & 56 960 & 60 809 &   \\
\bottomrule
\bottomrule
\end{tabular}
\end{table}
\begin{table}
\caption{Comparison of theoretical energy level calculations for Cn II and Cn III. The energy levels of this work are given by $E_{\text{CIPT}}$ with $g-$factors $g_{\text{CIPT}}$. Previous atomic calculations using relativistic coupled-cluster calculations ($E_{\text{RCC}}$) are presented for comparison. }
\begin{tabular}{l@{\hspace{0.5cm}}cr@{\hspace{0.5cm}}r@{\hspace{0.5cm}}r@{\hspace{0.5cm}}r@{\hspace{0.5cm}}r}
\toprule
\toprule
& State  & $E_{\text{CIPT}}$ & $g_{\text{CIPT}}$ &  $E_{\text{RCC}}$ \cite{Eliav_Cn_1995} \\
 \midrule
								 \multicolumn{5}{c}{Cn II}\\
$6d^9 7s^2$   &  $^2$D$_{5/2}$   &       0 &  1.20 &    \\
$6d^{10} 7s $ & $^2$S$_{1/2}$   &          11 037 & 2.00 & 12 905\\
$6d^9 7s^2$ & $^2$D$_{3/2}$   &      23 760&  0.80 & 25 326 \\
$6d^8 7s^2 7p$  & 1$_{7/2}^{\rm o}$ &        53 236 &  1.31 &     \\
\\
\midrule
								 \multicolumn{5}{c}{Cn III} \\
$6d^8 7s^2$  & 1$_4$    &   0 &  1.23      \\
$6d^8 7s^2$ & 1$_2$    &      681  &   1.08  & 374 \\
$6d^9 7s$   & 1$_3$   &       1 160  &  1.33  & 1 493 \\
$6d^{10}$    & $^1$S$_0$ &     8 521 &    0.00  & 6 411 \\
$6d^9 7s$       &  1$_1$ &      27 029 &   0.50 & 28 353 \\
$6d^8 7s  7p$ &1$_1^{\rm o}$ & 64 336 & 1.29 \\
\bottomrule
\bottomrule
\end{tabular}
\end{table}
 The CIPT calculations of Hg I were performed using the reference configurations $5d^{10}6s^2$ (even states) and, $5d^{9}6s^26p$ and $5d^{10}6s6p$ (odd states). For Hg II we used reference states $5d^{9}6s^2$ and $5d^{10}6s$ (even states) and, $5d^{8}6s^26p$, $5d^{9}6s6p$ and $5d^{10}6p$ (odd states). For Hg II we used reference states $5d^{8}6s^2$, $5d^{9}6s$ and $5d^{10}$ (even states) and, $5d^{7}6s^26p$, $5d^{8}6s6p$ and $5d^{9}6p$ (odd states). The same sets of reference configurations were used for the Cn I-III calculations with the appropriate principal quantum numbers.
\section{Electric Dipole transitions and isotope shifts} \label{sec:E1IS}
Along with the excitation spectrum we also calculated the electric dipole transition rates, $A_{\text{E1}}$, for allowed transitions to ground with transition frequencies $\omega <$ 45~000~cm$^{-1}$ which are presented in Table \ref{tab:E1Is}. Only these transitions are considered as they are the ones that can be measured with the current experimental spectroscopy methods for heavy elements \cite{Laatiaoui2016, Laatiaoui20161, Backe2015}. The maximum transition frequency currently accessible is $\omega \approx$ 40~000$^{-1}$\cite{LaatiaouiPC} so states up to $\omega = $45~000~cm$^{-1}$ are presented to account for the uncertainty in the calculations and future experimental advancements. These are some of the first spectroscopic properties to be measured in experiments and therefore theoretical predictions will aid future experiments.
The E1 transition rates are calculated using the formula 
\begin{align} \label{eq:AE1}
A_{E1} = \dfrac{4}{3}\left(\alpha \omega\right)^3\dfrac{ D_{E1}^2}{2J + 1}
\end{align}
where $J$ is the angular momentum of the upper state and $D_{E1}$ is the E1 transition amplitude. The E1 transition amplitudes are calculated using a self-consistent random-phase approximation (RPA) (see Refs. \cite{Dzuba2018, LDFDb2018} for more details). The accuracy of these calculations was discussed in Refs. \cite{LDFSg2019, LDFOg2018} by calculating the transition rates of light analogs and comparing to experiment. It was found that while the accuracy of the E1 rates was not on the same level as the energy spectrum calculations, they were in agreement to an order of magnitude. This is due to the $\omega^3$ proportionality in Eq. (\ref{eq:AE1}) which drastically decreases the accuracy of $A_{\text{E1}}$ for reasonably small deviations in accuracy for $\omega$ (energy levels). However, as these rates are primarily used to identify promising states for experimental measurements, this level of accuracy is sufficient. All possible strong optical E1 transitions for the neutral atoms and ions considered in Sections \ref{sec:DsRg} and \ref{sec:Cn} are presented in Table \ref{tab:E1Is}.\\
\linebreak
From Table \ref{tab:E1Is} we see there are several optically accessible states for Ds I compared to Pt I. However, few of these states have large transitions rates. The transitions with the largest rates are $^3$F$_4 \rightarrow$~2$_5^{\rm o}$, $^3$F$_4 \rightarrow$~4$_4^{\rm o}$ and $^3$F$_4 \rightarrow$~2$_4^{\rm o}$. For Ds II the promising transitions are $^4$F$_{9/2} \rightarrow$ 1$_{9/2}^{\rm o}$ and $^4$F$_{9/2} \rightarrow$ 2$_{9/2}^{\rm o}$.\\
\linebreak
Along with these strong E1 transitions we also calculate the isotope shift (IS) of the these energy levels. The IS is an important property as it is an indirect indicator of the effect of the nucleus on the atomic properties of the atoms. The IS can be used to find the difference in nuclear radius between two isotopes and, if the spectra of lighter neutron deficient isotopes is known,  predict the spectra of heavier, meta-stable neutron rich isotopes. This can be used to identify long sought after meta-stable super heavy nuclei in the spectra of astronomical data \cite{DFW17, Polukhina2012, Gopka2008, Fivet2007}. The effect of the IS is separated into two different mechanisms, the volume shifts which dominates in SHE\cite{Stacey1966} and the mass shift which is negligible in for heavy elements. Therefore, in this work we only consider the effect of the volume shift. Using the CIPT method, we calculate the excitation spectrum of the each isotope by varying the nuclear radius in the nuclear potential in the HF procedure described in Section \ref{sec:CIPT}. We present three different IS parameters based on different models of the IS. The first form of the IS is,
\begin{align} \label{eq:isoa}
\delta \nu &= E_{2} - E_{1} = a\left(A_{2}^{2\gamma/3} - A_{1}^{2\gamma/3}\right),
\end{align}
where $A_1$ and $A_2$ are atomic numbers for two isotopes ($A_2>A_1$), $E_1$ and $E_2$ are the excitation energy for  $A_1$ and $A_2$ respectively and $a$ is a parameter which should be calculated for each transition. This form is based on the approximation the isotope shift is dependent on $R_N^{2 \gamma}$ where  $\gamma = \sqrt{1-(Z\alpha)^2}$ and the large scale trend of nuclear radius $R_N \propto A^{1/3}$, see Refs. \cite{LDFSg2019, FGV2018} for more details. This form of the IS is convenient for isotopes with large differences in atomic number and therefore particularly useful for predicting the spectra of meta-stable isotopes from lighter isotopes synthesized in laboratories. However it should be noted that the large scale trend of nuclear radius and nuclear volume is not necessarily valid for SHEs due to the non-uniform density of the nucleus. This may leads to large deviations in the calculated IS \cite{Nazarewicz2018}. \\
\linebreak
The two last forms of the IS presented are related to the  root mean squared nuclear radius, $ R_{rms} = \sqrt{\left<r^{2}\right>}$ which is the nuclear charge radius of the nucleus and calculated using a Fermi distribution to model the nuclear density. A common form of isotope shift is the relation between the change of atomic frequency to the change of nuclear charge radius
\begin{align} \label{eq:isoF}
\delta \nu &= F\delta R_{rms}^2,
\end{align}
This formula (neglecting the mass shift) is convenient for extraction of the nuclear charge radius change from isotope shift measurements of nearby isotopes. The final form of IS we present was introduced in our previous work Ref. \cite{LDFSg2019}
\begin{align}\label{eq:isoFtilde}
\delta \nu = \tilde{F}\dfrac{R_{rms,A_2}^{2\gamma} - R_{rms,A_1}^{2\gamma}}{\text{fm}^{2\gamma}}
\end{align}
where $\tilde{F}$ is an IS parameter to be calculated for each transition. This form is valid for all isotope calculations and is based on the IS proportionality mentioned above, $\delta \nu \propto \delta R_{rms}^{2\gamma}$.\\
\linebreak
For the lighter isotope in the IS calculation, we calculated the spectra using $^{272}$Ds ($R_{rms,272}=5.8534 \ \text{fm}^{2}$) and $^{272}$Rg ($R_{rms,272}=5.8534 \ \text{fm}^{2}$). For the meta-stable isotope with $N=184$ we used $^{294}$Ds ($R_{rms,294}=6.039 \ \text{fm}^{2}$) and  $^{295}$Rg ($R_{rms,295}=6.0452 \ \text{fm}^{2}$). The isotope shift associated with the strong E1 transitions are presented in Table \ref{tab:E1Is}.
\begin{table*}
\caption{Strong electric dipole transition amplitudes, $D_{\text{E1}}$, transition rates, $A_{\text{E1}}$ from the ground state to the upper odd parity states of Ds I, Ds II, Rg I and  Rg II . Isotope shift parameters $a$, $F$ and $\tilde{F}$ between lighter, synthesized isotopes and theoretically metastable stable isotope with neutron number $N=184$ are also presented. \label{tab:E1Is} }
\begin{tabular}{c@{\hspace{0.5cm}}c@{\hspace{0.5cm}}r@{\hspace{0.5cm}}r@{\hspace{0.5cm}}r@{\hspace{0.5cm}}r@{\hspace{0.5cm}}r@{\hspace{0.5cm}}r@{\hspace{0.5cm}}r@{\hspace{0.5cm}}r}
\toprule
\toprule
Upper State & Energy (cm$^{-1}$) &   \parbox{1cm}{$D_{\text{E1}}$ \\ (a.u)} & \parbox{1cm}{$A_{\text{E1}}$ \\ { \small $(\times 10^{6} \ \text{s}^{-1})$ }} & \parbox{1cm}{$a  $ \\ (cm$^{-1}$)} & \parbox{1cm}{$F $ \\ ($\frac{\text{cm}^{-1}}{\text{fm}^{2}}$)} &   \multicolumn{1}{c}{\parbox{1cm}{$\tilde{F} $ \\ (cm$^{-1}$) }}\\
 \midrule
\multicolumn{7}{c}{Ds I } \\
 \midrule
 1$_4^{\rm o}$  &  21 812 & 0.318  & 0.236 & 64.5 & 5.07 & 37.8 \\
1$_5^{\rm o}$  &  24 958 &  0.00846 & .000205 & 65.7 & 5.17 & 38.6 \\
1$_3^{\rm o}$  &  32 645 & 0.125 & 0.156 & 50.6 & 3.98 & 29.7 \\
2$_4^{\rm o}$  &  34 919 & 1.132 & 12.3 & -183 & -14.4 & -108 \\
2$_3^{\rm o}$ &   36 404 & 0.0790  & 0.0873 & 64.5 & 5.07 & 37.8 \\
3$_4^{\rm o}$  &  39 814 & 0.464 & 3.06 & 85.9 & 6.76 & 50.5 \\
2$_5^{\rm o}$  &   40 173 &  2.06 & 50.9 & -120 & -9.45 & -70.5 \\
3$_5^{\rm o}$  & 42 682  &  0.631 & 5.72 &  -34.1 & -2.68 & -20.0  \\
3$_3^{\rm o}$ &  42 722   & 0.359  & 2.90 &  -77.1 & -6.07 & -45.3  \\
4$_4^{\rm o}$  & 43 915  & 1.08  & 22.2  & -27.8 & -2.19 & -16.3  \\
\\
\multicolumn{7}{c}{Ds II} \\
 \midrule
 1$_{7/2}^{\rm o}$ &      37 951 & 0.135 & 0.251 & 49.3 & 3.88 & 28.9 \\
 1$_{9/2}^{\rm o}$  &   39 010 & 0.949 & 10.8 & -206 & -16.2 & -121 \\
 2$_{9/2}^{\rm o}$  &      41 911 & 0.898 & 12.0 & 36.7 & 2.88 & 21.5 \\ 
 \\

 \multicolumn{7}{c}{Rg I} \\
  \midrule
 1$_{7/2}^{\rm o}$  &   28 224 & 0.115 &  0.0753 & 80.9 & 6.15 & 46.0\\
1$_{3/2}^{\rm o}$ &  32 677 & 0.487  & 4.20  & 58.1 & 4.42 & 33.0 \\
1$_{5/2}^{\rm o}$  & 34 398 & 0.374 & 1.92  & 56.9 & 4.33 & 32.3 \\
$^4$P$_{5/2}^{\rm o}$  & 42 709 & 0.932 & 22.9 & -235 & -17.9 & -134  \\
\\
\multicolumn{7}{c}{Rg II } \\
\midrule
1$_4^{\rm o}$ &   42 047 & 0.290 & 1.41 & 69.5 & 5.29 & 39.5 \\
\bottomrule
\bottomrule
\end{tabular}
\end{table*}
\section{Successive Ionization potentials} \label{sec:IP}
In this section we calculate the successive ionization potentials for Ds, Rg and Cn. Along with the strong dipole transitions, this ionization potential is one of the first atomic properties of elements to be measured. The ionization potential of elements also reveals details about the chemical and spectroscopic properties of the elements. In Table \ref{tab:IPs}  we present the ionic states and ionization potentials of Ds I, II, III, IV, and Rg, I, II, III, IV, V, and Cn I, II, III, IV, V, VI.
\begin{table*}
\caption{Successive ionization potential calculations using CIPT method  for SHEs Ds, Rg and Cn. Calculations for the lighter analogs are also presented for  comparison with experimental results. %The configurations presented are those for the ionic states and
 Energies marked with an asterisk (*) denote the theoretical calculations listed in the NIST database \cite{NIST_ASD}. Those values are quoted as having an uncertainty of 10 000- 15 000 (cm$^{-1}$). \label{tab:IPs}}
\begin{tabular}{l@{\hspace{0.3cm}}l@{\hspace{0.3cm}}l@{\hspace{0.3cm}}c@{\hspace{0.3cm}}r@{\hspace{0.3cm}}r @{\hspace{0.3cm}}r @{\hspace{0.3cm}}@{\hspace{0.3cm}} r@{\hspace{0.3cm}}l@{\hspace{0.3cm}}l@{\hspace{0.3cm}}l@{\hspace{0.3cm}}c @{\hspace{0.3cm}}r@{\hspace{0.3cm}}r@{\hspace{0.3cm}}r }
\toprule
\toprule
 &        &   \multicolumn{2}{c}{Ground State}  &     &       \multicolumn{3}{c}{IP (cm$^{-1}$)}  &  & & \multicolumn{2}{c}{Ground State} & & \multicolumn{2}{c}{IP (cm$^{-1}$)}  \\
 \cline{3-4}  \cline{6-8} \cline{11-12} \cline{14-15} \\
  & Ion & Config. & Term & & Expt. & CIPT & $\Delta$ & & Ion & Config. & $J$ & & CIPT & Other \\
  \midrule
Pt & I & $5d^96s$ &   $^3$D$_3$ & & 72 257.8 & 73 225 & -967 & Ds & I & $6d^8 7s^2$ & $^3$F$_4$  & & 81 933 & 89 984$^a$ \\
 & II & $5d^9$ &$^{2}$D$_{5/2}$  & & 149 723  & 150 026 & -303 &	&	 II & $6d^7 7s^2$ &$^{4}$F$_{9/2}$ &  & 141 108 \\
 & III & $5d^8$ &$^{3}$F$_{4}$   & & 234 000* & 245 806  & -11 806 & & III & $6d^6 7s^2$ &1$_2$ &  & 240 185   \\
 & IV & $5d^7$  &$^{4}$F$_{9/2}$  & & $347 000$* & 353 657 & -6 657 & & IV & $6d^67s$ &$^{6}$D$_{9/2}$ &  & 328 830 \\ \\
 Au & I & $5d^{10}6s$ & $^2$S$_{1/2}$ & & 74 409.11 & 75 776 & -1 367 & Rg & I & $6d^9 7s^2$ & $^2$D$_{5/2}$ & & 90 132 & 98 764$^a$, 95 748$^b$   \\
& II & $5d^{10}$ & $^1$S$_0$ & & 162 950 & 165 104 & -2 154 & & II &  $6d^8 7s^2$ &$^{3}$F$_4$ &  & 171 989   \\
& III & $5d^9$ &$^{2}$D$_{5/2}$   & & 242 000* & 260 197 & -18 197 & & III & $6d^7 7s^2$ & $^{4}$F$_{9/2}$ & & 250 503 \\  
 & IV & $5d^8$  &$^{3}$F$_{4}$  &  & 363 000* & 368 951& -5 951 & & IV & $6d^77s$ & $^5$F$_5$   & & 338 736 \\ 
 & V & $5d^7$ &$^{4}$F$_{9/2}$  & & 484 000* &  488 769 & -4 769 & & V & $6d^7$ &$^{4}$F$_{9/2}$   & & 439 861\\ \\
 Hg & I & $5d^{10} 6s^2$ & $^1$S$_0$  & &  84 184.15 & 84 782  & -598 & Cn & I & $6d^{10} 7s^2$ & $^1$S$_0$ & & 97 956 & 105 336$^a$, 91 569$^b$, 94 609$^c$ \\ 
 & II & $5d^{10}6s$ & $^2$S$_{1/2}$ &  & 151 284.4  & 152 120 & -836 & & II &  $6d^9 7s^2$ &$^2$D$_{5/2}$   & & 184 241 & 177 354$^b$, 177 281$^c$\\
 & III & $5d^{10}$ &$^1$S$_0$  & & 277 900* & 280 295 & -2 395  & & III & $6d^97s$ &$^3$D$_3$ &  & 260 665 & 263 098$^c$  \\
 & IV & $5d^{9}$ & $^{2}$D$_{5/2}$& & 391 600* & 386 525 & 5 075 & & IV & $6d^9$ &$^{2}$D$_{5/2}$ &  & 351 903 \\
 & V & $5d^{8}$ &$^{3}$F$_{4}$   & & 493 600* & 506 264 & -12 664 & & V & $6d^8$ &$^{3}$F$_{4}$ &  & 451 630  \\
 & VI & $5d^{7}$ &$^{4}$F$_{9/2}$  & & 618 000* & 636 714 & -18 714 & & VI & $6d^7$ &$^{4}$F$_{9/2}$ &  & 566 242 \\

\bottomrule
\bottomrule
\end{tabular}
\begin{flushleft}
$^a$ Ref. \cite{Dzuba2016}, $^b$ Ref. \cite{Hengele2012}, $^c$  Ref. \cite{Yu2007}
\end{flushleft}
\end{table*}
To calculate the IP of each element a new basis is constructed using the $V^{N-1}$ approximation for each successive ionization. The CIPT method is then used to calculate the energies of the ground state and an ionic state. In Table \ref{tab:IPs} we present the successive ionization potentials of the SHEs along with analogous calculations for lighter elements for comparison with experimental results. Comparing the CIPT IPs to the experimental values for the lighter elements we see good agreement between the results with discrepancy of  few percent where experimental results are available in Ref. \cite{NIST_ASD}.\\
% For the theoretical values (marked with an *) there is significantly worse agreement for the light analogs however these values have extremely large uncertainties.\\
\linebreak
The IP of some of the ionic species have been calculated previously and are included for reference in Table \ref{tab:IPs}. Comparing our results with previous calculations we see there is an agreement to within 10~000~cm$^{-1}$. In Ref. \cite{Yu2007} the IPs of all the neutral atoms and ions in group 12 were calculated using an MCDF method. When compared to experimental results these values are consistently 5-10\% lower than experimental values. 
% Victor This does not make sense. in Ref. \cite{Yu2007} values are lower than experiment. Your values are 10% lower than their values, so they are 20% wrong.
%This  agrees with our calculation of the IPs of Cn I, II, III which are 10\% lower too. 

The results of Ref. \cite{Dzuba2016} were  calculated by extrapolating a term in the Hamiltonian of the relativistic Hartree-Fock potential. This extrapolation was included to agree with those of lighter analogs. However as the first ionization of lighter elements is due to the removal of a $6s$ electron, the SHEs are ionized by first removing a $6d$ electron. Therefore, the extrapolation in Ref. \cite{Dzuba2016} may not be accurate, though the calculations agree with ours to about 10\% which was similar agreement found for lighter SHEs calculated in \cite{LDFSg2019} using the same method.\\
\section{Conclusion}
The improved calculation and understanding of atomic properties of SHEs is important in aiding future experiments on these elements. In this paper we calculated the low-lying atomic spectrum of Ds, Rg, Cn and their ions. Promising strong E1 transitions for future experimental measurement were calculated for these atoms and ions. The isotope shift parameters calculated will hopefully facilitate the detection of nuclei from the island of stability which has been long sought after. In this paper we also calculated the successive ionization potentials of the SHE. The ionization potential is one of the first measured properties of elements and therefore these calculations should aid in experimental studies. 
\bibliographystyle{apsrev4-1}
\bibliography{SHE_110_112}

\begin{thebibliography}{47}%
\makeatletter
\providecommand \@ifxundefined [1]{%
 \@ifx{#1\undefined}
}%
\providecommand \@ifnum [1]{%
 \ifnum #1\expandafter \@firstoftwo
 \else \expandafter \@secondoftwo
 \fi
}%
\providecommand \@ifx [1]{%
 \ifx #1\expandafter \@firstoftwo
 \else \expandafter \@secondoftwo
 \fi
}%
\providecommand \natexlab [1]{#1}%
\providecommand \enquote  [1]{``#1''}%
\providecommand \bibnamefont  [1]{#1}%
\providecommand \bibfnamefont [1]{#1}%
\providecommand \citenamefont [1]{#1}%
\providecommand \href@noop [0]{\@secondoftwo}%
\providecommand \href [0]{\begingroup \@sanitize@url \@href}%
\providecommand \@href[1]{\@@startlink{#1}\@@href}%
\providecommand \@@href[1]{\endgroup#1\@@endlink}%
\providecommand \@sanitize@url [0]{\catcode `\\12\catcode `\$12\catcode
  `\&12\catcode `\#12\catcode `\^12\catcode `\_12\catcode `\%12\relax}%
\providecommand \@@startlink[1]{}%
\providecommand \@@endlink[0]{}%
\providecommand \url  [0]{\begingroup\@sanitize@url \@url }%
\providecommand \@url [1]{\endgroup\@href {#1}{\urlprefix }}%
\providecommand \urlprefix  [0]{URL }%
\providecommand \Eprint [0]{\href }%
\providecommand \doibase [0]{http://dx.doi.org/}%
\providecommand \selectlanguage [0]{\@gobble}%
\providecommand \bibinfo  [0]{\@secondoftwo}%
\providecommand \bibfield  [0]{\@secondoftwo}%
\providecommand \translation [1]{[#1]}%
\providecommand \BibitemOpen [0]{}%
\providecommand \bibitemStop [0]{}%
\providecommand \bibitemNoStop [0]{.\EOS\space}%
\providecommand \EOS [0]{\spacefactor3000\relax}%
\providecommand \BibitemShut  [1]{\csname bibitem#1\endcsname}%
\let\auto@bib@innerbib\@empty
%</preamble>
\bibitem [{\citenamefont {Giuliani}\ \emph {et~al.}(2019)\citenamefont
  {Giuliani}, \citenamefont {Matheson}, \citenamefont {Nazarewicz},
  \citenamefont {Olsen}, \citenamefont {Reinhard}, \citenamefont {Sadhukhan},
  \citenamefont {Schuetrumpf}, \citenamefont {Schunck},\ and\ \citenamefont
  {Schwerdtfeger}}]{Giuliani2019}%
  \BibitemOpen
  \bibfield  {author} {\bibinfo {author} {\bibfnamefont {S.~A.}\ \bibnamefont
  {Giuliani}}, \bibinfo {author} {\bibfnamefont {Z.}~\bibnamefont {Matheson}},
  \bibinfo {author} {\bibfnamefont {W.}~\bibnamefont {Nazarewicz}}, \bibinfo
  {author} {\bibfnamefont {E.}~\bibnamefont {Olsen}}, \bibinfo {author}
  {\bibfnamefont {P.~G.}\ \bibnamefont {Reinhard}}, \bibinfo {author}
  {\bibfnamefont {J.}~\bibnamefont {Sadhukhan}}, \bibinfo {author}
  {\bibfnamefont {B.}~\bibnamefont {Schuetrumpf}}, \bibinfo {author}
  {\bibfnamefont {N.}~\bibnamefont {Schunck}}, \ and\ \bibinfo {author}
  {\bibfnamefont {P.}~\bibnamefont {Schwerdtfeger}},\ }\href@noop {} {\bibfield
   {journal} {\bibinfo  {journal} {Rev. Mod. Phys.}\ }\textbf {\bibinfo
  {volume} {91}},\ \bibinfo {pages} {011001} (\bibinfo {year}
  {2019})}\BibitemShut {NoStop}%
\bibitem [{\citenamefont {Sch{\"a}del}(2015)}]{schadel2015}%
  \BibitemOpen
  \bibfield  {author} {\bibinfo {author} {\bibfnamefont {M.}~\bibnamefont
  {Sch{\"a}del}},\ }\href@noop {} {\bibfield  {journal} {\bibinfo  {journal}
  {Philosophical Transactions of the Royal Society of London A: Mathematical,
  Physical and Engineering Sciences}\ }\textbf {\bibinfo {volume} {373}},\
  \bibinfo {pages} {20140191} (\bibinfo {year} {2015})}\BibitemShut {NoStop}%
\bibitem [{\citenamefont {Hamilton}\ \emph {et~al.}(2013)\citenamefont
  {Hamilton}, \citenamefont {Hofmann},\ and\ \citenamefont
  {Oganessian}}]{HHO2013}%
  \BibitemOpen
  \bibfield  {author} {\bibinfo {author} {\bibfnamefont {J.~H.}\ \bibnamefont
  {Hamilton}}, \bibinfo {author} {\bibfnamefont {S.}~\bibnamefont {Hofmann}}, \
  and\ \bibinfo {author} {\bibfnamefont {Y.~T.}\ \bibnamefont {Oganessian}},\
  }\href@noop {} {\bibfield  {journal} {\bibinfo  {journal} {Annu. Rev. Nucl.
  Part. Sci.}\ }\textbf {\bibinfo {volume} {63}},\ \bibinfo {pages} {383}
  (\bibinfo {year} {2013})}\BibitemShut {NoStop}%
\bibitem [{\citenamefont {Laatiaoui}\ \emph {et~al.}(2016)\citenamefont
  {Laatiaoui}, \citenamefont {Lauth}, \citenamefont {Backe}, \citenamefont
  {Block},\ and\ \citenamefont {Ackermann}}]{Laatiaoui2016}%
  \BibitemOpen
  \bibfield  {author} {\bibinfo {author} {\bibfnamefont {M.}~\bibnamefont
  {Laatiaoui}}, \bibinfo {author} {\bibfnamefont {W.}~\bibnamefont {Lauth}},
  \bibinfo {author} {\bibfnamefont {H.}~\bibnamefont {Backe}}, \bibinfo
  {author} {\bibfnamefont {M.}~\bibnamefont {Block}}, \ and\ \bibinfo {author}
  {\bibfnamefont {D.}~\bibnamefont {Ackermann}},\ }\href@noop {} {\bibfield
  {journal} {\bibinfo  {journal} {Nature}\ }\textbf {\bibinfo {volume} {538}},\
  \bibinfo {pages} {495} (\bibinfo {year} {2016})}\BibitemShut {NoStop}%
\bibitem [{\citenamefont {Chhetri}\ \emph {et~al.}(2018)\citenamefont
  {Chhetri}, \citenamefont {Ackermann}, \citenamefont {Backe}, \citenamefont
  {Block}, \citenamefont {Cheal}, \citenamefont {Droese}, \citenamefont
  {Dullmann}, \citenamefont {Even}, \citenamefont {Ferrer}, \citenamefont
  {Giacoppo}, \citenamefont {Gotz}, \citenamefont {Hessberger}, \citenamefont
  {Huyse}, \citenamefont {Kaleja}, \citenamefont {Khuyagbaatar}, \citenamefont
  {Kunz}, \citenamefont {Laatiaoui}, \citenamefont {Lautenschlager},
  \citenamefont {Lauth}, \citenamefont {Lecesne}, \citenamefont {Lens},
  \citenamefont {MinayaRamirez}, \citenamefont {Mistry}, \citenamefont
  {Raeder}, \citenamefont {VanDuppen}, \citenamefont {Walther}, \citenamefont
  {Yakushev},\ and\ \citenamefont {Zhang}}]{Chhetri2018}%
  \BibitemOpen
  \bibfield  {author} {\bibinfo {author} {\bibfnamefont {P.}~\bibnamefont
  {Chhetri}}, \bibinfo {author} {\bibfnamefont {D.}~\bibnamefont {Ackermann}},
  \bibinfo {author} {\bibfnamefont {H.}~\bibnamefont {Backe}}, \bibinfo
  {author} {\bibfnamefont {M.}~\bibnamefont {Block}}, \bibinfo {author}
  {\bibfnamefont {B.}~\bibnamefont {Cheal}}, \bibinfo {author} {\bibfnamefont
  {C.}~\bibnamefont {Droese}}, \bibinfo {author} {\bibfnamefont {C.~E.}\
  \bibnamefont {Dullmann}}, \bibinfo {author} {\bibfnamefont {J.}~\bibnamefont
  {Even}}, \bibinfo {author} {\bibfnamefont {R.}~\bibnamefont {Ferrer}},
  \bibinfo {author} {\bibfnamefont {F.}~\bibnamefont {Giacoppo}}, \bibinfo
  {author} {\bibfnamefont {S.}~\bibnamefont {Gotz}}, \bibinfo {author}
  {\bibfnamefont {F.~P.}\ \bibnamefont {Hessberger}}, \bibinfo {author}
  {\bibfnamefont {M.}~\bibnamefont {Huyse}}, \bibinfo {author} {\bibfnamefont
  {O.}~\bibnamefont {Kaleja}}, \bibinfo {author} {\bibfnamefont
  {J.}~\bibnamefont {Khuyagbaatar}}, \bibinfo {author} {\bibfnamefont
  {P.}~\bibnamefont {Kunz}}, \bibinfo {author} {\bibfnamefont {M.}~\bibnamefont
  {Laatiaoui}}, \bibinfo {author} {\bibfnamefont {F.}~\bibnamefont
  {Lautenschlager}}, \bibinfo {author} {\bibfnamefont {W.}~\bibnamefont
  {Lauth}}, \bibinfo {author} {\bibfnamefont {N.}~\bibnamefont {Lecesne}},
  \bibinfo {author} {\bibfnamefont {L.}~\bibnamefont {Lens}}, \bibinfo {author}
  {\bibfnamefont {E.}~\bibnamefont {MinayaRamirez}}, \bibinfo {author}
  {\bibfnamefont {A.~K.}\ \bibnamefont {Mistry}}, \bibinfo {author}
  {\bibfnamefont {S.}~\bibnamefont {Raeder}}, \bibinfo {author} {\bibfnamefont
  {P.}~\bibnamefont {VanDuppen}}, \bibinfo {author} {\bibfnamefont
  {T.}~\bibnamefont {Walther}}, \bibinfo {author} {\bibfnamefont
  {A.}~\bibnamefont {Yakushev}}, \ and\ \bibinfo {author} {\bibfnamefont
  {Z.}~\bibnamefont {Zhang}},\ }\href@noop {} {\bibfield  {journal} {\bibinfo
  {journal} {Phys. Rev. Lett.}\ }\textbf {\bibinfo {volume} {120}},\ \bibinfo
  {pages} {263003} (\bibinfo {year} {2018})}\BibitemShut {NoStop}%
\bibitem [{\citenamefont {Sato}\ \emph {et~al.}(2015)\citenamefont {Sato},
  \citenamefont {Asai}, \citenamefont {Borschevsky}, \citenamefont {Stora},
  \citenamefont {Sato}, \citenamefont {Kaneya},\ and\ \citenamefont
  {Tsukada}}]{SAB15}%
  \BibitemOpen
  \bibfield  {author} {\bibinfo {author} {\bibfnamefont {T.~K.}\ \bibnamefont
  {Sato}}, \bibinfo {author} {\bibfnamefont {M.}~\bibnamefont {Asai}}, \bibinfo
  {author} {\bibfnamefont {A.}~\bibnamefont {Borschevsky}}, \bibinfo {author}
  {\bibfnamefont {T.}~\bibnamefont {Stora}}, \bibinfo {author} {\bibfnamefont
  {N.}~\bibnamefont {Sato}}, \bibinfo {author} {\bibfnamefont {Y.}~\bibnamefont
  {Kaneya}}, \ and\ \bibinfo {author} {\bibfnamefont {K.}~\bibnamefont
  {Tsukada}},\ }\href@noop {} {\bibfield  {journal} {\bibinfo  {journal}
  {Nature}\ }\textbf {\bibinfo {volume} {520}},\ \bibinfo {pages} {209}
  (\bibinfo {year} {2015})}\BibitemShut {NoStop}%
\bibitem [{\citenamefont {Eliav}\ \emph {et~al.}(2015)\citenamefont {Eliav},
  \citenamefont {Fritzsche},\ and\ \citenamefont {Kaldor}}]{eliav2015}%
  \BibitemOpen
  \bibfield  {author} {\bibinfo {author} {\bibfnamefont {E.}~\bibnamefont
  {Eliav}}, \bibinfo {author} {\bibfnamefont {S.}~\bibnamefont {Fritzsche}}, \
  and\ \bibinfo {author} {\bibfnamefont {U.}~\bibnamefont {Kaldor}},\
  }\href@noop {} {\bibfield  {journal} {\bibinfo  {journal} {Nucl. Phys. A}\
  }\textbf {\bibinfo {volume} {944}},\ \bibinfo {pages} {518} (\bibinfo {year}
  {2015})}\BibitemShut {NoStop}%
\bibitem [{\citenamefont {Pershina}(2015)}]{pershina2015}%
  \BibitemOpen
  \bibfield  {author} {\bibinfo {author} {\bibfnamefont {V.}~\bibnamefont
  {Pershina}},\ }\href@noop {} {\bibfield  {journal} {\bibinfo  {journal}
  {Nucl. Phys. A}\ }\textbf {\bibinfo {volume} {944}},\ \bibinfo {pages} {578}
  (\bibinfo {year} {2015})}\BibitemShut {NoStop}%
\bibitem [{\citenamefont {Dzuba}\ \emph
  {et~al.}(2017{\natexlab{a}})\citenamefont {Dzuba}, \citenamefont {Berengut},
  \citenamefont {Harabati},\ and\ \citenamefont {Flambaum}}]{DBHF2017}%
  \BibitemOpen
  \bibfield  {author} {\bibinfo {author} {\bibfnamefont {V.~A.}\ \bibnamefont
  {Dzuba}}, \bibinfo {author} {\bibfnamefont {J.~C.}\ \bibnamefont {Berengut}},
  \bibinfo {author} {\bibfnamefont {C.}~\bibnamefont {Harabati}}, \ and\
  \bibinfo {author} {\bibfnamefont {V.~V.}\ \bibnamefont {Flambaum}},\
  }\href@noop {} {\bibfield  {journal} {\bibinfo  {journal} {Phys. Rev. A}\
  }\textbf {\bibinfo {volume} {95}},\ \bibinfo {pages} {012503} (\bibinfo
  {year} {2017}{\natexlab{a}})}\BibitemShut {NoStop}%
\bibitem [{\citenamefont {Lackenby}\ \emph
  {et~al.}(2018{\natexlab{a}})\citenamefont {Lackenby}, \citenamefont {Dzuba},\
  and\ \citenamefont {Flambaum}}]{LDFDb2018}%
  \BibitemOpen
  \bibfield  {author} {\bibinfo {author} {\bibfnamefont {B.~G.~C.}\
  \bibnamefont {Lackenby}}, \bibinfo {author} {\bibfnamefont {V.~A.}\
  \bibnamefont {Dzuba}}, \ and\ \bibinfo {author} {\bibfnamefont {V.~V.}\
  \bibnamefont {Flambaum}},\ }\href@noop {} {\bibfield  {journal} {\bibinfo
  {journal} {Phys. Rev. A}\ }\textbf {\bibinfo {volume} {98}},\ \bibinfo
  {pages} {022518} (\bibinfo {year} {2018}{\natexlab{a}})}\BibitemShut
  {NoStop}%
\bibitem [{\citenamefont {Lackenby}\ \emph {et~al.}(2019)\citenamefont
  {Lackenby}, \citenamefont {Dzuba},\ and\ \citenamefont
  {Flambaum}}]{LDFSg2019}%
  \BibitemOpen
  \bibfield  {author} {\bibinfo {author} {\bibfnamefont {B.~G.~C.}\
  \bibnamefont {Lackenby}}, \bibinfo {author} {\bibfnamefont {V.~A.}\
  \bibnamefont {Dzuba}}, \ and\ \bibinfo {author} {\bibfnamefont {V.~V.}\
  \bibnamefont {Flambaum}},\ }\href@noop {} {\bibfield  {journal} {\bibinfo
  {journal} {Phys. Rev. A}\ }\textbf {\bibinfo {volume} {99}},\ \bibinfo
  {pages} {042509} (\bibinfo {year} {2019})}\BibitemShut {NoStop}%
\bibitem [{\citenamefont {Lackenby}\ \emph
  {et~al.}(2018{\natexlab{b}})\citenamefont {Lackenby}, \citenamefont {Dzuba},\
  and\ \citenamefont {Flambaum}}]{LDFOg2018}%
  \BibitemOpen
  \bibfield  {author} {\bibinfo {author} {\bibfnamefont {B.~G.~C.}\
  \bibnamefont {Lackenby}}, \bibinfo {author} {\bibfnamefont {V.~A.}\
  \bibnamefont {Dzuba}}, \ and\ \bibinfo {author} {\bibfnamefont {V.~V.}\
  \bibnamefont {Flambaum}},\ }\href@noop {} {\bibfield  {journal} {\bibinfo
  {journal} {Phys. Rev. A}\ }\textbf {\bibinfo {volume} {98}},\ \bibinfo
  {pages} {042512} (\bibinfo {year} {2018}{\natexlab{b}})}\BibitemShut
  {NoStop}%
\bibitem [{\citenamefont {Goriely}\ \emph {et~al.}(2017)\citenamefont
  {Goriely}, \citenamefont {Bauswein},\ and\ \citenamefont
  {Janka}}]{Goriely2011}%
  \BibitemOpen
  \bibfield  {author} {\bibinfo {author} {\bibfnamefont {S.}~\bibnamefont
  {Goriely}}, \bibinfo {author} {\bibfnamefont {A.}~\bibnamefont {Bauswein}}, \
  and\ \bibinfo {author} {\bibfnamefont {H.-T.}\ \bibnamefont {Janka}},\
  }\href@noop {} {\bibfield  {journal} {\bibinfo  {journal} {Astrophys. J.
  Lett.}\ }\textbf {\bibinfo {volume} {738}},\ \bibinfo {pages} {L32} (\bibinfo
  {year} {2017})}\BibitemShut {NoStop}%
\bibitem [{\citenamefont {Fuller}\ \emph {et~al.}(2017)\citenamefont {Fuller},
  \citenamefont {Kusenko},\ and\ \citenamefont {Takhistov}}]{Fuller2017}%
  \BibitemOpen
  \bibfield  {author} {\bibinfo {author} {\bibfnamefont {G.~M.}\ \bibnamefont
  {Fuller}}, \bibinfo {author} {\bibfnamefont {A.}~\bibnamefont {Kusenko}}, \
  and\ \bibinfo {author} {\bibfnamefont {V.}~\bibnamefont {Takhistov}},\
  }\href@noop {} {\bibfield  {journal} {\bibinfo  {journal} {Phys. Rev. Lett.}\
  }\textbf {\bibinfo {volume} {119}},\ \bibinfo {pages} {061101} (\bibinfo
  {year} {2017})}\BibitemShut {NoStop}%
\bibitem [{\citenamefont {Frebel}\ and\ \citenamefont
  {Beers}(2018)}]{Friebel2018}%
  \BibitemOpen
  \bibfield  {author} {\bibinfo {author} {\bibfnamefont {A.}~\bibnamefont
  {Frebel}}\ and\ \bibinfo {author} {\bibfnamefont {T.~C.}\ \bibnamefont
  {Beers}},\ }\href@noop {} {\bibfield  {journal} {\bibinfo  {journal} {Phys.
  Today}\ }\textbf {\bibinfo {volume} {71}},\ \bibinfo {pages} {30} (\bibinfo
  {year} {2018})}\BibitemShut {NoStop}%
\bibitem [{\citenamefont {Schuetrumpf}\ \emph {et~al.}(2015)\citenamefont
  {Schuetrumpf}, \citenamefont {Klatt}, \citenamefont {Iida}, \citenamefont
  {Schr\"{o}der-Turk}, \citenamefont {Maruhn}, \citenamefont {Mecke},\ and\
  \citenamefont {Reinhard}}]{Schuetrumpf2015}%
  \BibitemOpen
  \bibfield  {author} {\bibinfo {author} {\bibfnamefont {B.}~\bibnamefont
  {Schuetrumpf}}, \bibinfo {author} {\bibfnamefont {M.~A.}\ \bibnamefont
  {Klatt}}, \bibinfo {author} {\bibfnamefont {K.}~\bibnamefont {Iida}},
  \bibinfo {author} {\bibfnamefont {G.~E.}\ \bibnamefont {Schr\"{o}der-Turk}},
  \bibinfo {author} {\bibfnamefont {J.~A.}\ \bibnamefont {Maruhn}}, \bibinfo
  {author} {\bibfnamefont {K.}~\bibnamefont {Mecke}}, \ and\ \bibinfo {author}
  {\bibfnamefont {P.~G.}\ \bibnamefont {Reinhard}},\ }\href@noop {} {\bibfield
  {journal} {\bibinfo  {journal} {Phys. Rev. C}\ }\textbf {\bibinfo {volume}
  {91}},\ \bibinfo {pages} {025801} (\bibinfo {year} {2015})}\BibitemShut
  {NoStop}%
\bibitem [{\citenamefont {Dzuba}\ \emph
  {et~al.}(2017{\natexlab{b}})\citenamefont {Dzuba}, \citenamefont {Flambaum},\
  and\ \citenamefont {Webb}}]{DFW17}%
  \BibitemOpen
  \bibfield  {author} {\bibinfo {author} {\bibfnamefont {V.~A.}\ \bibnamefont
  {Dzuba}}, \bibinfo {author} {\bibfnamefont {V.~V.}\ \bibnamefont {Flambaum}},
  \ and\ \bibinfo {author} {\bibfnamefont {J.~K.}\ \bibnamefont {Webb}},\
  }\href@noop {} {\bibfield  {journal} {\bibinfo  {journal} {Phys. Rev. A}\
  }\textbf {\bibinfo {volume} {95}},\ \bibinfo {pages} {062515} (\bibinfo
  {year} {2017}{\natexlab{b}})}\BibitemShut {NoStop}%
\bibitem [{\citenamefont {Li}\ \emph {et~al.}(2007)\citenamefont {Li},
  \citenamefont {Dong}, \citenamefont {Yu}, \citenamefont {Ding}, \citenamefont
  {Fritzsche},\ and\ \citenamefont {Fricke}}]{Li2007}%
  \BibitemOpen
  \bibfield  {author} {\bibinfo {author} {\bibfnamefont {J.~G.}\ \bibnamefont
  {Li}}, \bibinfo {author} {\bibfnamefont {C.~Z.}\ \bibnamefont {Dong}},
  \bibinfo {author} {\bibfnamefont {Y.~J.}\ \bibnamefont {Yu}}, \bibinfo
  {author} {\bibfnamefont {X.~B.}\ \bibnamefont {Ding}}, \bibinfo {author}
  {\bibfnamefont {S.}~\bibnamefont {Fritzsche}}, \ and\ \bibinfo {author}
  {\bibfnamefont {B.}~\bibnamefont {Fricke}},\ }\href@noop {} {\bibfield
  {journal} {\bibinfo  {journal} {Eur. Phys. J. D}\ }\textbf {\bibinfo {volume}
  {44}},\ \bibinfo {pages} {51} (\bibinfo {year} {2007})}\BibitemShut {NoStop}%
\bibitem [{\citenamefont {Yu}\ \emph {et~al.}(2007)\citenamefont {Yu},
  \citenamefont {Li}, \citenamefont {Dong}, \citenamefont {Ding}, \citenamefont
  {Fritzsche},\ and\ \citenamefont {Fricke}}]{Yu2007}%
  \BibitemOpen
  \bibfield  {author} {\bibinfo {author} {\bibfnamefont {Y.~J.}\ \bibnamefont
  {Yu}}, \bibinfo {author} {\bibfnamefont {J.~G.}\ \bibnamefont {Li}}, \bibinfo
  {author} {\bibfnamefont {C.~Z.}\ \bibnamefont {Dong}}, \bibinfo {author}
  {\bibfnamefont {X.~B.}\ \bibnamefont {Ding}}, \bibinfo {author}
  {\bibfnamefont {S.}~\bibnamefont {Fritzsche}}, \ and\ \bibinfo {author}
  {\bibfnamefont {B.}~\bibnamefont {Fricke}},\ }\href@noop {} {\bibfield
  {journal} {\bibinfo  {journal} {Eur. Phys. J. D}\ }\textbf {\bibinfo {volume}
  {44}},\ \bibinfo {pages} {51} (\bibinfo {year} {2007})}\BibitemShut {NoStop}%
\bibitem [{\citenamefont {Hangele}\ \emph {et~al.}(2012)\citenamefont
  {Hangele}, \citenamefont {Dolg}, \citenamefont {Hanrath}, \citenamefont
  {Cao},\ and\ \citenamefont {Schwerdtfeger}}]{Hengele2012}%
  \BibitemOpen
  \bibfield  {author} {\bibinfo {author} {\bibfnamefont {T.}~\bibnamefont
  {Hangele}}, \bibinfo {author} {\bibfnamefont {M.}~\bibnamefont {Dolg}},
  \bibinfo {author} {\bibfnamefont {M.}~\bibnamefont {Hanrath}}, \bibinfo
  {author} {\bibfnamefont {X.}~\bibnamefont {Cao}}, \ and\ \bibinfo {author}
  {\bibfnamefont {P.}~\bibnamefont {Schwerdtfeger}},\ }\href@noop {} {\bibfield
   {journal} {\bibinfo  {journal} {J. Chem. Phys.}\ }\textbf {\bibinfo {volume}
  {136}},\ \bibinfo {pages} {214105} (\bibinfo {year} {2012})}\BibitemShut
  {NoStop}%
\bibitem [{\citenamefont {Dinh}\ \emph {et~al.}(2008)\citenamefont {Dinh},
  \citenamefont {Dzuba},\ and\ \citenamefont {Flambaum}}]{Dinh2008}%
  \BibitemOpen
  \bibfield  {author} {\bibinfo {author} {\bibfnamefont {T.~H.}\ \bibnamefont
  {Dinh}}, \bibinfo {author} {\bibfnamefont {V.~A.}\ \bibnamefont {Dzuba}}, \
  and\ \bibinfo {author} {\bibfnamefont {V.~V.}\ \bibnamefont {Flambaum}},\
  }\href {\doibase 10.1103/PhysRevA.78.062502} {\bibfield  {journal} {\bibinfo
  {journal} {Phys. Rev. A}\ }\textbf {\bibinfo {volume} {78}},\ \bibinfo
  {pages} {062502} (\bibinfo {year} {2008})}\BibitemShut {NoStop}%
\bibitem [{\citenamefont {Eliav}\ \emph {et~al.}(1995)\citenamefont {Eliav},
  \citenamefont {Kaldor},\ and\ \citenamefont {Ishikawa}}]{Eliav_Cn_1995}%
  \BibitemOpen
  \bibfield  {author} {\bibinfo {author} {\bibfnamefont {E.}~\bibnamefont
  {Eliav}}, \bibinfo {author} {\bibfnamefont {U.}~\bibnamefont {Kaldor}}, \
  and\ \bibinfo {author} {\bibfnamefont {Y.}~\bibnamefont {Ishikawa}},\
  }\href@noop {} {\bibfield  {journal} {\bibinfo  {journal} {Phys. Rev. A}\
  }\textbf {\bibinfo {volume} {52}},\ \bibinfo {pages} {2765} (\bibinfo {year}
  {1995})}\BibitemShut {NoStop}%
\bibitem [{\citenamefont {Dzuba}\ \emph
  {et~al.}(2018{\natexlab{a}})\citenamefont {Dzuba}, \citenamefont {Flambaum},\
  and\ \citenamefont {Kozlov}}]{FCI}%
  \BibitemOpen
  \bibfield  {author} {\bibinfo {author} {\bibfnamefont {V.~A.}\ \bibnamefont
  {Dzuba}}, \bibinfo {author} {\bibfnamefont {V.~V.}\ \bibnamefont {Flambaum}},
  \ and\ \bibinfo {author} {\bibfnamefont {M.~G.}\ \bibnamefont {Kozlov}},\
  }\href@noop {} {\  (\bibinfo {year} {2018}{\natexlab{a}})},\ \bibinfo {note}
  {arXiv:1812.11480}\BibitemShut {NoStop}%
\bibitem [{\citenamefont {Kelly}(1964)}]{Kelly1964}%
  \BibitemOpen
  \bibfield  {author} {\bibinfo {author} {\bibfnamefont {H.~P.}\ \bibnamefont
  {Kelly}},\ }\href@noop {} {\bibfield  {journal} {\bibinfo  {journal} {Phys.
  Rev.}\ }\textbf {\bibinfo {volume} {136}},\ \bibinfo {pages} {3B} (\bibinfo
  {year} {1964})}\BibitemShut {NoStop}%
\bibitem [{\citenamefont {Dzuba}(2005)}]{Dzuba2005}%
  \BibitemOpen
  \bibfield  {author} {\bibinfo {author} {\bibfnamefont {V.~A.}\ \bibnamefont
  {Dzuba}},\ }\href@noop {} {\bibfield  {journal} {\bibinfo  {journal} {Phys.
  Rev. A}\ }\textbf {\bibinfo {volume} {71}},\ \bibinfo {pages} {032512}
  (\bibinfo {year} {2005})}\BibitemShut {NoStop}%
\bibitem [{\citenamefont {Breit}(1929)}]{Breit1929}%
  \BibitemOpen
  \bibfield  {author} {\bibinfo {author} {\bibfnamefont {G.}~\bibnamefont
  {Breit}},\ }\href@noop {} {\bibfield  {journal} {\bibinfo  {journal} {Phys.
  Rev.}\ }\textbf {\bibinfo {volume} {34}},\ \bibinfo {pages} {4} (\bibinfo
  {year} {1929})}\BibitemShut {NoStop}%
\bibitem [{\citenamefont {Mann}\ and\ \citenamefont
  {Johnson}(1971)}]{Mann1971}%
  \BibitemOpen
  \bibfield  {author} {\bibinfo {author} {\bibfnamefont {J.~B.}\ \bibnamefont
  {Mann}}\ and\ \bibinfo {author} {\bibfnamefont {W.~R.}\ \bibnamefont
  {Johnson}},\ }\href@noop {} {\bibfield  {journal} {\bibinfo  {journal} {Phys.
  Rev. A}\ }\textbf {\bibinfo {volume} {4}},\ \bibinfo {pages} {1} (\bibinfo
  {year} {1971})}\BibitemShut {NoStop}%
\bibitem [{\citenamefont {Flambaum}\ and\ \citenamefont
  {Ginges}(2005)}]{FG2005}%
  \BibitemOpen
  \bibfield  {author} {\bibinfo {author} {\bibfnamefont {V.~V.}\ \bibnamefont
  {Flambaum}}\ and\ \bibinfo {author} {\bibfnamefont {J.~S.~M.}\ \bibnamefont
  {Ginges}},\ }\href {\doibase 10.1103/PhysRevA.72.052115} {\bibfield
  {journal} {\bibinfo  {journal} {Phys. Rev. A}\ }\textbf {\bibinfo {volume}
  {72}},\ \bibinfo {pages} {052115} (\bibinfo {year} {2005})}\BibitemShut
  {NoStop}%
\bibitem [{\citenamefont {Dzuba}\ and\ \citenamefont
  {Flambaum}(2016)}]{FF113-115}%
  \BibitemOpen
  \bibfield  {author} {\bibinfo {author} {\bibfnamefont {V.~A.}\ \bibnamefont
  {Dzuba}}\ and\ \bibinfo {author} {\bibfnamefont {V.~V.}\ \bibnamefont
  {Flambaum}},\ }\href@noop {} {\bibfield  {journal} {\bibinfo  {journal}
  {Hyperfine Interactions}\ }\textbf {\bibinfo {volume} {237}},\ \bibinfo
  {pages} {160} (\bibinfo {year} {2016})}\BibitemShut {NoStop}%
\bibitem [{\citenamefont {Hofmann}\ \emph
  {et~al.}(1995{\natexlab{a}})\citenamefont {Hofmann}, \citenamefont {Ninov},
  \citenamefont {He{\ss}berger}, \citenamefont {Armbruster}, \citenamefont
  {Folger}, \citenamefont {M{\"u}nzenberg}, \citenamefont {Sch{\"o}tt},
  \citenamefont {Popeko}, \citenamefont {Yeremin}, \citenamefont {Andreyev},
  \citenamefont {Saro}, \citenamefont {Janik},\ and\ \citenamefont
  {Leino}}]{Ds1994}%
  \BibitemOpen
  \bibfield  {author} {\bibinfo {author} {\bibfnamefont {S.}~\bibnamefont
  {Hofmann}}, \bibinfo {author} {\bibfnamefont {V.}~\bibnamefont {Ninov}},
  \bibinfo {author} {\bibfnamefont {F.~P.}\ \bibnamefont {He{\ss}berger}},
  \bibinfo {author} {\bibfnamefont {P.}~\bibnamefont {Armbruster}}, \bibinfo
  {author} {\bibfnamefont {H.}~\bibnamefont {Folger}}, \bibinfo {author}
  {\bibfnamefont {G.}~\bibnamefont {M{\"u}nzenberg}}, \bibinfo {author}
  {\bibfnamefont {H.~J.}\ \bibnamefont {Sch{\"o}tt}}, \bibinfo {author}
  {\bibfnamefont {A.~G.}\ \bibnamefont {Popeko}}, \bibinfo {author}
  {\bibfnamefont {A.~V.}\ \bibnamefont {Yeremin}}, \bibinfo {author}
  {\bibfnamefont {A.~N.}\ \bibnamefont {Andreyev}}, \bibinfo {author}
  {\bibfnamefont {S.}~\bibnamefont {Saro}}, \bibinfo {author} {\bibfnamefont
  {R.}~\bibnamefont {Janik}}, \ and\ \bibinfo {author} {\bibfnamefont
  {M.}~\bibnamefont {Leino}},\ }\href@noop {} {\bibfield  {journal} {\bibinfo
  {journal} {Z. Phys. A}\ }\textbf {\bibinfo {volume} {350}},\ \bibinfo {pages}
  {277} (\bibinfo {year} {1995}{\natexlab{a}})}\BibitemShut {NoStop}%
\bibitem [{\citenamefont {Hofmann}\ \emph
  {et~al.}(1995{\natexlab{b}})\citenamefont {Hofmann}, \citenamefont {Ninov},
  \citenamefont {He{\ss}berger}, \citenamefont {Armbruster}, \citenamefont
  {Folger}, \citenamefont {M{\"u}nzenberg}, \citenamefont {Sch{\"o}tt},
  \citenamefont {Popeko}, \citenamefont {Yeremin}, \citenamefont {Andreyev},
  \citenamefont {Saro}, \citenamefont {Janik},\ and\ \citenamefont
  {Leino}}]{Rg1994}%
  \BibitemOpen
  \bibfield  {author} {\bibinfo {author} {\bibfnamefont {S.}~\bibnamefont
  {Hofmann}}, \bibinfo {author} {\bibfnamefont {V.}~\bibnamefont {Ninov}},
  \bibinfo {author} {\bibfnamefont {F.~P.}\ \bibnamefont {He{\ss}berger}},
  \bibinfo {author} {\bibfnamefont {P.}~\bibnamefont {Armbruster}}, \bibinfo
  {author} {\bibfnamefont {H.}~\bibnamefont {Folger}}, \bibinfo {author}
  {\bibfnamefont {G.}~\bibnamefont {M{\"u}nzenberg}}, \bibinfo {author}
  {\bibfnamefont {H.~J.}\ \bibnamefont {Sch{\"o}tt}}, \bibinfo {author}
  {\bibfnamefont {A.~G.}\ \bibnamefont {Popeko}}, \bibinfo {author}
  {\bibfnamefont {A.~V.}\ \bibnamefont {Yeremin}}, \bibinfo {author}
  {\bibfnamefont {A.~N.}\ \bibnamefont {Andreyev}}, \bibinfo {author}
  {\bibfnamefont {S.}~\bibnamefont {Saro}}, \bibinfo {author} {\bibfnamefont
  {R.}~\bibnamefont {Janik}}, \ and\ \bibinfo {author} {\bibfnamefont
  {M.}~\bibnamefont {Leino}},\ }\href@noop {} {\bibfield  {journal} {\bibinfo
  {journal} {Z. Phys. A}\ }\textbf {\bibinfo {volume} {350}},\ \bibinfo {pages}
  {281} (\bibinfo {year} {1995}{\natexlab{b}})}\BibitemShut {NoStop}%
\bibitem [{\citenamefont {Karol}\ \emph {et~al.}(2001)\citenamefont {Karol},
  \citenamefont {Nakahara}, \citenamefont {Petley},\ and\ \citenamefont
  {Vogt}}]{DsIUPAC}%
  \BibitemOpen
  \bibfield  {author} {\bibinfo {author} {\bibfnamefont {P.~J.}\ \bibnamefont
  {Karol}}, \bibinfo {author} {\bibfnamefont {H.}~\bibnamefont {Nakahara}},
  \bibinfo {author} {\bibfnamefont {B.~W.}\ \bibnamefont {Petley}}, \ and\
  \bibinfo {author} {\bibfnamefont {E.}~\bibnamefont {Vogt}},\ }\href@noop {}
  {\bibfield  {journal} {\bibinfo  {journal} {Pure App. Chem.}\ }\textbf
  {\bibinfo {volume} {73}},\ \bibinfo {pages} {959} (\bibinfo {year}
  {2001})}\BibitemShut {NoStop}%
\bibitem [{\citenamefont {Eliav}\ \emph {et~al.}(1994)\citenamefont {Eliav},
  \citenamefont {Kaldor}, \citenamefont {Schwerdtfeger}, \citenamefont {Hess},\
  and\ \citenamefont {Ishikawa}}]{Eliav_Rg_1994}%
  \BibitemOpen
  \bibfield  {author} {\bibinfo {author} {\bibfnamefont {E.}~\bibnamefont
  {Eliav}}, \bibinfo {author} {\bibfnamefont {U.}~\bibnamefont {Kaldor}},
  \bibinfo {author} {\bibfnamefont {P.}~\bibnamefont {Schwerdtfeger}}, \bibinfo
  {author} {\bibfnamefont {B.~A.}\ \bibnamefont {Hess}}, \ and\ \bibinfo
  {author} {\bibfnamefont {Y.}~\bibnamefont {Ishikawa}},\ }\href@noop {}
  {\bibfield  {journal} {\bibinfo  {journal} {Phys. Rev. Lett.}\ }\textbf
  {\bibinfo {volume} {73}},\ \bibinfo {pages} {3203} (\bibinfo {year}
  {1994})}\BibitemShut {NoStop}%
\bibitem [{\citenamefont {Kramida}\ \emph {et~al.}(2018)\citenamefont
  {Kramida}, \citenamefont {{Yu.~Ralchenko}}, \citenamefont {Reader},\ and\
  \citenamefont {{and NIST ASD Team}}}]{NIST_ASD}%
  \BibitemOpen
  \bibfield  {author} {\bibinfo {author} {\bibfnamefont {A.}~\bibnamefont
  {Kramida}}, \bibinfo {author} {\bibnamefont {{Yu.~Ralchenko}}}, \bibinfo
  {author} {\bibfnamefont {J.}~\bibnamefont {Reader}}, \ and\ \bibinfo {author}
  {\bibnamefont {{and NIST ASD Team}}},\ }\href@noop {} {}\bibinfo
  {howpublished} {{NIST Atomic Spectra Database (ver. 5.6.1), [Online].
  Available: {\tt{https://physics.nist.gov/asd}} [2019, August 14]. National
  Institute of Standards and Technology, Gaithersburg, MD.}} (\bibinfo {year}
  {2018})\BibitemShut {NoStop}%
\bibitem [{\citenamefont {Hofmann}\ \emph {et~al.}(1996)\citenamefont
  {Hofmann}, \citenamefont {Ninov}, \citenamefont {Hessberger}, \citenamefont
  {Armbruster}, \citenamefont {Folger}, \citenamefont {M\"{u}nzenberg},
  \citenamefont {Sch\"{o}tt}, \citenamefont {G.}, \citenamefont {Yeremin},
  \citenamefont {Saro}, \citenamefont {Janik},\ and\ \citenamefont
  {Leino}}]{Hofmann1996}%
  \BibitemOpen
  \bibfield  {author} {\bibinfo {author} {\bibfnamefont {S.}~\bibnamefont
  {Hofmann}}, \bibinfo {author} {\bibfnamefont {I.~V.}\ \bibnamefont {Ninov}},
  \bibinfo {author} {\bibfnamefont {F.~P.}\ \bibnamefont {Hessberger}},
  \bibinfo {author} {\bibfnamefont {P.}~\bibnamefont {Armbruster}}, \bibinfo
  {author} {\bibfnamefont {H.}~\bibnamefont {Folger}}, \bibinfo {author}
  {\bibfnamefont {G.}~\bibnamefont {M\"{u}nzenberg}}, \bibinfo {author}
  {\bibfnamefont {H.~J.}\ \bibnamefont {Sch\"{o}tt}}, \bibinfo {author}
  {\bibfnamefont {P.~A.}\ \bibnamefont {G.}}, \bibinfo {author} {\bibfnamefont
  {A.~V.}\ \bibnamefont {Yeremin}}, \bibinfo {author} {\bibfnamefont
  {S.}~\bibnamefont {Saro}}, \bibinfo {author} {\bibfnamefont {R.}~\bibnamefont
  {Janik}}, \ and\ \bibinfo {author} {\bibfnamefont {M.}~\bibnamefont
  {Leino}},\ }\href@noop {} {\bibfield  {journal} {\bibinfo  {journal} {Z.
  Phyisik A}\ }\textbf {\bibinfo {volume} {354}},\ \bibinfo {pages} {229}
  (\bibinfo {year} {1996})}\BibitemShut {NoStop}%
\bibitem [{\citenamefont {Eichler}\ \emph {et~al.}(2007)\citenamefont
  {Eichler}, \citenamefont {Aksenov}, \citenamefont {Belozerov}, \citenamefont
  {Bozhikov}, \citenamefont {Chepigin}, \citenamefont {Dmitriev},\ and\
  \citenamefont {Dressler}}]{Eichler2007}%
  \BibitemOpen
  \bibfield  {author} {\bibinfo {author} {\bibfnamefont {R.}~\bibnamefont
  {Eichler}}, \bibinfo {author} {\bibfnamefont {N.~V.}\ \bibnamefont
  {Aksenov}}, \bibinfo {author} {\bibfnamefont {A.~V.}\ \bibnamefont
  {Belozerov}}, \bibinfo {author} {\bibfnamefont {G.~A.}\ \bibnamefont
  {Bozhikov}}, \bibinfo {author} {\bibfnamefont {V.~I.}\ \bibnamefont
  {Chepigin}}, \bibinfo {author} {\bibfnamefont {S.~N.}\ \bibnamefont
  {Dmitriev}}, \ and\ \bibinfo {author} {\bibfnamefont {R.}~\bibnamefont
  {Dressler}},\ }\href@noop {} {\bibfield  {journal} {\bibinfo  {journal}
  {Nature}\ }\textbf {\bibinfo {volume} {447}},\ \bibinfo {pages} {72}
  (\bibinfo {year} {2007})}\BibitemShut {NoStop}%
\bibitem [{\citenamefont {Laatiaoui}(2016)}]{Laatiaoui20161}%
  \BibitemOpen
  \bibfield  {author} {\bibinfo {author} {\bibfnamefont {M.}~\bibnamefont
  {Laatiaoui}},\ }\href@noop {} {\bibfield  {journal} {\bibinfo  {journal} {EPJ
  Web Conf.}\ }\textbf {\bibinfo {volume} {131}},\ \bibinfo {pages} {05002}
  (\bibinfo {year} {2016})}\BibitemShut {NoStop}%
\bibitem [{\citenamefont {Backe}\ \emph {et~al.}(2015)\citenamefont {Backe},
  \citenamefont {Lauth}, \citenamefont {Block},\ and\ \citenamefont
  {Laatiaoui}}]{Backe2015}%
  \BibitemOpen
  \bibfield  {author} {\bibinfo {author} {\bibfnamefont {H.}~\bibnamefont
  {Backe}}, \bibinfo {author} {\bibfnamefont {W.}~\bibnamefont {Lauth}},
  \bibinfo {author} {\bibfnamefont {M.}~\bibnamefont {Block}}, \ and\ \bibinfo
  {author} {\bibfnamefont {M.}~\bibnamefont {Laatiaoui}},\ }\href@noop {}
  {\bibfield  {journal} {\bibinfo  {journal} {Nucl. Phys. A}\ }\textbf
  {\bibinfo {volume} {944}},\ \bibinfo {pages} {492} (\bibinfo {year}
  {2015})}\BibitemShut {NoStop}%
\bibitem [{\citenamefont {Laatiaoui}()}]{LaatiaouiPC}%
  \BibitemOpen
  \bibfield  {author} {\bibinfo {author} {\bibfnamefont {M.}~\bibnamefont
  {Laatiaoui}},\ }\href@noop {} {}\bibinfo {note} {Private
  Communication.}\BibitemShut {Stop}%
\bibitem [{\citenamefont {Dzuba}\ \emph
  {et~al.}(2018{\natexlab{b}})\citenamefont {Dzuba}, \citenamefont {Flambaum},\
  and\ \citenamefont {Schiller}}]{Dzuba2018}%
  \BibitemOpen
  \bibfield  {author} {\bibinfo {author} {\bibfnamefont {V.~A.}\ \bibnamefont
  {Dzuba}}, \bibinfo {author} {\bibfnamefont {V.~V.}\ \bibnamefont {Flambaum}},
  \ and\ \bibinfo {author} {\bibfnamefont {S.}~\bibnamefont {Schiller}},\
  }\href@noop {} {\bibfield  {journal} {\bibinfo  {journal} {Phys. Rev. A}\
  }\textbf {\bibinfo {volume} {98}},\ \bibinfo {pages} {022501} (\bibinfo
  {year} {2018}{\natexlab{b}})}\BibitemShut {NoStop}%
\bibitem [{\citenamefont {Polukhina}(2012)}]{Polukhina2012}%
  \BibitemOpen
  \bibfield  {author} {\bibinfo {author} {\bibfnamefont {N.~G.}\ \bibnamefont
  {Polukhina}},\ }\href@noop {} {\bibfield  {journal} {\bibinfo  {journal}
  {Phys.-Usp.}\ }\textbf {\bibinfo {volume} {55}},\ \bibinfo {pages} {614}
  (\bibinfo {year} {2012})}\BibitemShut {NoStop}%
\bibitem [{\citenamefont {Gopka}\ \emph {et~al.}(2008)\citenamefont {Gopka},
  \citenamefont {Yushchenko}, \citenamefont {Yushchenko}, \citenamefont
  {Panov},\ and\ \citenamefont {Kim}}]{Gopka2008}%
  \BibitemOpen
  \bibfield  {author} {\bibinfo {author} {\bibfnamefont {V.~F.}\ \bibnamefont
  {Gopka}}, \bibinfo {author} {\bibfnamefont {A.~V.}\ \bibnamefont
  {Yushchenko}}, \bibinfo {author} {\bibfnamefont {V.~A.}\ \bibnamefont
  {Yushchenko}}, \bibinfo {author} {\bibfnamefont {I.~V.}\ \bibnamefont
  {Panov}}, \ and\ \bibinfo {author} {\bibfnamefont {C.}~\bibnamefont {Kim}},\
  }\href@noop {} {\bibfield  {journal} {\bibinfo  {journal} {Kinematics Phys.
  Celestial Bodies}\ }\textbf {\bibinfo {volume} {24}},\ \bibinfo {pages} {89}
  (\bibinfo {year} {2008})}\BibitemShut {NoStop}%
\bibitem [{\citenamefont {Fivet}\ \emph {et~al.}(2007)\citenamefont {Fivet},
  \citenamefont {Quinet}, \citenamefont {Bi\'{e}mont}, \citenamefont
  {Jorissen}, \citenamefont {Yushchenko},\ and\ \citenamefont
  {Van~Eck}}]{Fivet2007}%
  \BibitemOpen
  \bibfield  {author} {\bibinfo {author} {\bibfnamefont {V.}~\bibnamefont
  {Fivet}}, \bibinfo {author} {\bibfnamefont {P.}~\bibnamefont {Quinet}},
  \bibinfo {author} {\bibfnamefont {E.}~\bibnamefont {Bi\'{e}mont}}, \bibinfo
  {author} {\bibfnamefont {A.}~\bibnamefont {Jorissen}}, \bibinfo {author}
  {\bibfnamefont {A.~V.}\ \bibnamefont {Yushchenko}}, \ and\ \bibinfo {author}
  {\bibfnamefont {S.}~\bibnamefont {Van~Eck}},\ }\href@noop {} {\bibfield
  {journal} {\bibinfo  {journal} {Mom. Not. R. Astron. Soc.}\ }\textbf
  {\bibinfo {volume} {380}},\ \bibinfo {pages} {781} (\bibinfo {year}
  {2007})}\BibitemShut {NoStop}%
\bibitem [{\citenamefont {Stacey}(1966)}]{Stacey1966}%
  \BibitemOpen
  \bibfield  {author} {\bibinfo {author} {\bibfnamefont {D.~N.}\ \bibnamefont
  {Stacey}},\ }\href@noop {} {\bibfield  {journal} {\bibinfo  {journal} {Rep.
  Prog. Phys.}\ }\textbf {\bibinfo {volume} {29}},\ \bibinfo {pages} {171}
  (\bibinfo {year} {1966})}\BibitemShut {NoStop}%
\bibitem [{\citenamefont {Flambaum}\ \emph {et~al.}(2018)\citenamefont
  {Flambaum}, \citenamefont {Geddes},\ and\ \citenamefont
  {Viatkina}}]{FGV2018}%
  \BibitemOpen
  \bibfield  {author} {\bibinfo {author} {\bibfnamefont {V.~V.}\ \bibnamefont
  {Flambaum}}, \bibinfo {author} {\bibfnamefont {A.~J.}\ \bibnamefont
  {Geddes}}, \ and\ \bibinfo {author} {\bibfnamefont {A.~V.}\ \bibnamefont
  {Viatkina}},\ }\href@noop {} {\bibfield  {journal} {\bibinfo  {journal}
  {Phys. Rev. A}\ }\textbf {\bibinfo {volume} {97}},\ \bibinfo {pages} {032510}
  (\bibinfo {year} {2018})}\BibitemShut {NoStop}%
\bibitem [{\citenamefont {Nazarewicz}(2018)}]{Nazarewicz2018}%
  \BibitemOpen
  \bibfield  {author} {\bibinfo {author} {\bibfnamefont {W.}~\bibnamefont
  {Nazarewicz}},\ }\href@noop {} {\bibfield  {journal} {\bibinfo  {journal}
  {Nat. Phys.}\ }\textbf {\bibinfo {volume} {14}},\ \bibinfo {pages} {537}
  (\bibinfo {year} {2018})}\BibitemShut {NoStop}%
\bibitem [{\citenamefont {Dzuba}(2016)}]{Dzuba2016}%
  \BibitemOpen
  \bibfield  {author} {\bibinfo {author} {\bibfnamefont {V.~A.}\ \bibnamefont
  {Dzuba}},\ }\href@noop {} {\bibfield  {journal} {\bibinfo  {journal} {Phys.
  Rev. A}\ }\textbf {\bibinfo {volume} {93}},\ \bibinfo {pages} {032519}
  (\bibinfo {year} {2016})}\BibitemShut {NoStop}%
\end{thebibliography}%
\end{document}